\documentclass[pra,nofootinbib,superscriptaddress,showpacs,twocolumn,longbibliography]{revtex4-2}

\usepackage{graphicx} 
\usepackage[colorlinks]{hyperref}
\usepackage{amsmath,amssymb, physics} 
\usepackage{tikz}
\usepackage{tikz-network}
\usepackage{xifthen}
\usepackage{xargs}

\newcommandx{\fineq}[5][1=-.8ex,2=1,3=1,5=0]{\begin{tikzpicture}[baseline={([yshift=#1]current  bounding  box.center)}, scale = #2, every node/.style={scale = #3},rotate around={#5:(0,0)},every node/.style={transform shape}]
		#4
	\end{tikzpicture}
}

\tikzset{
    cross/.pic = {
    \draw[very thick, rotate = 45] (-#1,0) -- (#1,0);
    \draw[very thick, rotate = 45] (0,-#1) -- (0, #1);
    }
}

\definecolor{bertinired}{RGB}{232,102,102}
\definecolor{bertiniblue}{RGB}{101,147,245}
\definecolor{bertinigreyblue}{RGB}{166,218,149}
\definecolor{bertinigreyred}{RGB}{232,102,102}
\definecolor{bertinivioletc}{RGB}{45,130,60}
\definecolor{bertinigreen}{RGB}{166,218,149}
\definecolor{bertiniorange}{RGB}{255, 116, 23}
\definecolor{OliveGreen}{RGB}{85,107,47}
\definecolor{NavyBlue}{RGB}{0,0,128}
\definecolor{bertiniviolet}{RGB}{210,145,178}
\definecolor{bertinigrey1}{RGB}{98,98,98}
\definecolor{bertinigrey2}{RGB}{211,211,211}
\definecolor{bertinigrey3}{RGB}{192,192,192}
\definecolor{bertinigrey4}{RGB}{169,169,169}

\newcommandx{\tikzdiagup}{
	\tikz {\draw[thick] (0,0)--(0.15,0.15); \draw (0,0) rectangle (0.15,0.15);}
}

\newcommandx{\gatecross}[1][1=0.5]{
	\pgfmathparse{#1/2.0}
	\let\x\pgfmathresult
	\draw[thick] (-\x,-\x) -- (\x,\x);
	\draw[thick] (\x,-\x) -- (-\x,\x);
}

\newcommandx{\gatecrossthin}[1][1=0.5]{
	\pgfmathparse{#1/2.0}
	\let\x\pgfmathresult
	\draw[] (-\x,-\x) -- (\x,\x);
	\draw[] (\x,-\x) -- (-\x,\x);
}

\newcommandx{\gatesqu}[2][1=0.25,2=]{
	\pgfmathparse{#1/2.0}
	\let\x\pgfmathresult
	\ifthenelse{\equal{#2}{}}{
		\draw[thick, fill=white, rounded corners=2pt] (-\x,\x) rectangle (\x,-\x);
	}{
		\draw[thick, fill=#2, rounded corners=2pt] (-\x,\x) rectangle (\x,-\x);
	}
}

\newcommandx{\gatesquthin}[2][1=0.25,2=]{
	\pgfmathparse{#1/2.0}
	\let\x\pgfmathresult
	\ifthenelse{\equal{#2}{}}{
		\draw[fill=white, rounded corners=2pt] (-\x,\x) rectangle (\x,-\x);
	}{
		\draw[fill=#2, rounded corners=2pt] (-\x,\x) rectangle (\x,-\x);
	}
}

\newcommandx{\gatemark}[2][1=0.075,2=tr]{
	\pgfmathparse{#1}
	\let\l\pgfmathresult
	\ifthenelse{\equal{#2}{topleft}}{
		\draw[thick] (-\l, 0.97*\l ) -- (\l, 0.97*\l) ;
	}{}
	\ifthenelse{\equal{#2}{topright}}{
		\draw[thick] (0.97*\l, \l) -- (0.97*\l, -\l);
	}{}

	\ifthenelse{\equal{#2}{bottomleft}}{
		\draw[thick] (0,-\l) -- ++(-\l,0) --++ (0,\l);
	}{}
	\ifthenelse{\equal{#2}{bottomright}}{
		\draw[thick] (0,-\l) -- ++(\l,0) --++ (0,\l);
	}{}
	
}

\newcommandx{\gatemarkthin}[2][1=0.075,2=tr]{
	\pgfmathparse{#1}
	\let\l\pgfmathresult
	\ifthenelse{\equal{#2}{topleft}}{
		\draw[] (-\l, 0.97*\l ) -- (\l, 0.97*\l) ;
	}{}
	\ifthenelse{\equal{#2}{topright}}{
		\draw[] (0.97*\l, \l) -- (0.97*\l, -\l);
	}{}

	\ifthenelse{\equal{#2}{bottomleft}}{
		\draw[] (0,-\l) -- ++(-\l,0) --++ (0,\l);
	}{}
	\ifthenelse{\equal{#2}{bottomright}}{
		\draw[] (0,-\l) -- ++(\l,0) --++ (0,\l);
	}{}
	
}

\newcommandx{\roundgate}[5][1=0,2=0,3=1,4=topright,5=white]{
	\pgfmathparse{#3}
	\let\l\pgfmathresult
	\begin{scope}[shift={(#1,#2)}]
		\gatecross[\l]
		
		\pgfmathparse{\l/ 8.0*5}

		\let\s\pgfmathresult
		\gatesqu[\s][#5]
		
		\pgfmathparse{\l*0.15}
		\let\m\pgfmathresult
		\gatemark[\m][#4]
	\end{scope}
}

\newcommandx{\roundgatethin}[5][1=0,2=0,3=1,4=topright,5=white]{
	\pgfmathparse{#3}
	\let\l\pgfmathresult
	\begin{scope}[shift={(#1,#2)}]
		\gatecrossthin[\l]
		
		\pgfmathparse{\l/ 8.0*5}

		\let\s\pgfmathresult
		\gatesquthin[\s][#5]
		
		\pgfmathparse{\l*0.15}
		\let\m\pgfmathresult
		\gatemarkthin[\m][#4]
	\end{scope}
}

\newcommandx{\wcirc}[2]{\begin{scope}
		\draw[fill=white] (#1,#2) circle (0.15);	\end{scope}} 
\newcommandx{\wcircc}[2]{\begin{scope}
		\draw[fill=white] (#1,#2) circle (0.13);	\end{scope}} 

\newcommandx{\wsqr}[2]{\begin{scope}
		\draw[fill=white,shift={(#1,#2)}] (-.13,.13) rectangle (.13,-.13);	\end{scope}} 
\newcommandx{\wsqrr}[2]{\begin{scope}
		\draw[fill=white,shift={(#1,#2)}] (-.11,.11) rectangle (.11,-.11);	\end{scope}} 
\newcommandx{\bcirc}[2]{\begin{scope}
		\draw[fill=black] (#1,#2) circle (0.15);	\end{scope}} 
\newcommandx{\thetastate}[4][1=0,2=0,3=1,4=]{
	\pgfmathparse{#3/2}
	\let\l\pgfmathresult
	\pgfmathparse{\l*0.15}
	\let\m\pgfmathresult
	\begin{scope}[shift={(#1,#2)}]
		\draw[thick] (0,0)--(\l,\l);
		\draw[thick] (0,0)--(-\l,\l);
		\ifthenelse{\equal{#4}{}}{
			\draw[fill=white] (0,0) circle (0.15);
		}{
			\draw[thick, fill=#4] (0,0) circle (0.15);
		}
	\end{scope}
}

\newcommandx{\thetastateflipped}[4][1=0,2=0,3=1,4=]{
	\pgfmathparse{#3/2}
	\let\l\pgfmathresult
	\pgfmathparse{\l*0.15}
	\let\m\pgfmathresult
	\begin{scope}[shift={(#1,#2)}]
		\draw[thick] (0,0)--(\l,-\l);
		\draw[thick] (0,0)--(-\l,-\l);
		\ifthenelse{\equal{#4}{}}{
			\draw[fill=white] (0,0) circle (0.15);
		}{
			\draw[thick, fill=#4] (0,0) circle (0.15);
		}
	\end{scope}
}

\newcommandx{\vertgate}[5][1=0,2=0,3=4,4=bertiniorange,5=topright]
{
	\begin{scope}[shift={(#1,#2)}]
		\ifthenelse{\equal{#3}{1}}{
			\roundgate[0][0][1][#5][#4]
		}{
			\foreach \n[evaluate=\n as \y using {2*\n-2}] in {1,...,#3}{
				\roundgate[0][\y][1][#5][#4]
			}
		}
	\end{scope}
}

\newcommandx{\tsfmatVthin}[8][1=0,2=0,3=l,4=4,5=tr,6=init,7=bertiniorange,8=topright]{
	\begin{scope}[shift={(#1,#2)}]
		\ifthenelse{\equal{#3}{l}}{
			\pgfmathsetmacro{\flag}{0}
		}{
			\pgfmathsetmacro{\flag}{1}
		}
		
		\foreach \y[evaluate=\y as \x using {mod(\y+\flag,2)}] in {1,...,#4}{
			\roundgatethin[\x][\y][1][#8][#7]
		}
		\ifthenelse{\equal{#5}{tr}}{
			\foreach \y[evaluate=\y as \x using {mod(\y+\flag,2)}] in {#4}{
				\draw [fill=white] (\x-0.5,\y+0.5) circle (0.15);
				\draw [fill=white] (\x+0.5,\y+0.5) circle (0.15);
			}
		}{}
		\ifthenelse{\equal{#6}{init}}{
			\thetastate[\flag][0][1][#7]
		}{}
	\end{scope}
}

\newcommandx{\tsfmatV}[8][1=0,2=0,3=l,4=4,5=tr,6=init,7=bertiniorange,8=topright]{
	\begin{scope}[shift={(#1,#2)}]
		\ifthenelse{\equal{#3}{l}}{
			\pgfmathsetmacro{\flag}{0}
		}{
			\pgfmathsetmacro{\flag}{1}
		}
		
		\foreach \y[evaluate=\y as \x using {mod(\y+\flag,2)}] in {1,...,#4}{
			\roundgate[\x][\y][1][#8][#7]
		}
		\ifthenelse{\equal{#5}{tr}}{
			\foreach \y[evaluate=\y as \x using {mod(\y+\flag,2)}] in {#4}{
				\draw [fill=white] (\x-0.5,\y+0.5) circle (0.15);
				\draw [fill=white] (\x+0.5,\y+0.5) circle (0.15);
			}
		}{}
		\ifthenelse{\equal{#6}{init}}{
			\thetastate[\flag][0][1][#7]
		}{}
	\end{scope}
}

\newcommandx{\leftriangle}[5][1=0,2=0,3=4,4=bertiniorange,5=topright]{
	\begin{scope}[shift={(#1,#2)}]
		\pgfmathsetmacro{\t}{#3}
		\pgfmathsetmacro{\steps}{ceil(\t/2)}
		\foreach \i[evaluate=\i as \x using -\t+2*\i-1, evaluate=\i as \ylim using \t-2*\i+2] in {1,...,\steps}{
			\foreach \y[evaluate=\y as \thisx using {\x+\y-1}] in {1,...,\ylim}{
				\roundgate[\thisx][\y][1][#5][#4]
			}
		}
	\end{scope}
}

\newcommandx{\rightriangle}[5][1=0,2=0,3=4,4=bertiniorange,5=topright]{
	\begin{scope}[shift={(#1,#2)}]
		\pgfmathsetmacro{\t}{#3}
		\pgfmathsetmacro{\steps}{ceil(\t/2)}
		\foreach \i[evaluate=\i as \x using -\t+2*\i-1, evaluate=\i as \ylim using \t-2*\i+2] in {1,...,\steps}{
			\foreach \y[evaluate=\y as \thisx using {-\x-\y+1}] in {1,...,\ylim}{
				\roundgate[\thisx][\y][1][#5][#4]
			}
		}
	\end{scope}
}

\newcommandx{\eigenVL}[8][1=0,2=0,3=l,4=5,5=tr,6=init,7=bertiniorange,8=topright]{
	\begin{scope}[shift={(#1,#2)}]
		\pgfmathsetmacro{\t}{#4}
		\leftriangle[0][0][\t][#7][#8]
		
		\ifthenelse{\equal{#6}{init}}{
			\drawinitstate[0][0][l][\t][#7]
		}{}
		
		\ifthenelse{\equal{#5}{tr}}{
			\draw[fill=white] \foreach \x in {0,...,\t} {(\x-0.5-\t,0.5+\x) circle (0.15)};
			\ifthenelse{\equal{#3}{r}}{
				\draw[fill=white] (0.5,\t+0.5) circle (0.15);
			}{}
		}{}
		\ifthenelse{\equal{#5}{parttr}}{
			\draw[fill=white] \foreach \x in {0,...,\t} {(\x-0.5-\t,0.5+\x) circle (0.15)};
		}{}
	\end{scope}
}

\newcommandx{\eigenVR}[8][1=0,2=0,3=l,4=5,5=tr,6=init,7=bertiniorange,8=topright]{
	\begin{scope}[shift={(#1,#2)}]
		\pgfmathsetmacro{\t}{#4}
		\rightriangle[0][0][\t][#7][#8]
		
		\ifthenelse{\equal{#6}{init}}{
			\drawinitstate[0][0][r][\t][#7]
		}{}
		
		\ifthenelse{\equal{#5}{tr}}{
			\draw[fill=white] \foreach \x in {0,...,\t}{(-\x+0.5+\t,0.5+\x) circle (0.15)};
			\ifthenelse{\equal{#3}{l}}{
				\draw[fill=white] (-0.5,\t+0.5) circle (0.15);
			}{}
		}{}
	\end{scope}
}

\newcommandx{\tra}[2][1]{\underset{#1}{\text{tr}}\left[#2\right]}

\newcommandx{\tsfmatDgate}[7][1=0,2=0,3=l,4=4,5=tr,6=bertiniorange,7=topright]
{
	\begin{scope}[shift={(#1,#2)}]
		\ifthenelse{\equal{#3}{l}}{
			\pgfmathsetmacro{\flag}{-1}
		}{
			\pgfmathsetmacro{\flag}{1}
		}
		\pgfmathsetmacro{\t}{#4}
		\foreach \i[evaluate=\i as \x using {\flag*\i}, evaluate=\i as \y using \i] in {1,...,\t}{
			\roundgate[\x][\y][1][#7][#6]
		}
		
		\ifthenelse{\equal{#5}{tr}}{
			\foreach \i[evaluate=\i as \x using {\flag*\i}, evaluate=\i as \y using \i] in {\t}{
				\draw [fill=white] (\x-0.5,\y+0.5) circle (0.15);
				\draw [fill=white] (\x+0.5,\y+0.5) circle (0.15);
			}  
		}{}
	\end{scope}
	
}

\newcommandx{\tsfmatD}[8][1=0,2=0,3=l,4=4,5=tr,6=init,7=bertiniorange,8=topright]{
	\begin{scope}[shift={(#1,#2)}]
		\ifthenelse{\equal{#6}{init}}{
			\thetastate[0][0][1][#7]
		}{}
		
		\ifthenelse{\equal{#3}{l}}{
			\pgfmathsetmacro{\flag}{-1}
		}{
			\pgfmathsetmacro{\flag}{1}
		}
		
		\pgfmathsetmacro{\t}{#4}
		\foreach \i[evaluate=\i as \x using {\flag*\i}, evaluate=\i as \y using \i] in {1,...,\t}{
			\roundgate[\x][\y][1][#8][#7]
		}
		
		\ifthenelse{\equal{#5}{tr}}{
			\foreach \i[evaluate=\i as \x using {\flag*\i}, evaluate=\i as \y using \i] in {\t}{
				\draw [fill=white] (\x-0.5,\y+0.5) circle (0.15);
				\draw [fill=white] (\x+0.5,\y+0.5) circle (0.15);
			}  
		}
		\ifthenelse{\equal{#5}{parttr}}{
			\foreach \i[evaluate=\i as \x using {\flag*\i}, evaluate=\i as \y using \i] in {\t}{
				\draw [fill=white] (\x+0.5*\flag,\y+0.5) circle (0.15);
			}  
		}
		{}
	\end{scope}
}

\newcommandx{\drawinitstate}[5][1=0,2=0,3=l,4=4,5=bertiniorange]{
	\pgfmathsetmacro{\t}{#4}
	\begin{scope}[shift={(#1,#2)}]
		\pgfmathsetmacro{\steps}{ceil((\t-1)/2)}
		\ifthenelse{\equal{#3}{l}}{
			\foreach \i[evaluate=\i as \x using -\t+2*\i] in {0,...,\steps}{
				\thetastate[\x][0][1][#5]
			}
		}{
			\foreach \i[evaluate=\i as \x using -\t+2*\i] in {0,...,\steps}{      
				\thetastate[-\x][0][1][#5]
			}
		}
	\end{scope}
}

\newcommandx{\drawinitstateflipped}[5][1=0,2=0,3=l,4=4,5=bertiniorange]{
	\pgfmathsetmacro{\t}{#4}
	\begin{scope}[shift={(#1,#2)}]
		\pgfmathsetmacro{\steps}{ceil((\t-1)/2)}
		\ifthenelse{\equal{#3}{l}}{
			\foreach \i[evaluate=\i as \x using -\t+2*\i] in {0,...,\steps}{
				\thetastateflipped[\x][0][1][#5]
			}
		}{
			\foreach \i[evaluate=\i as \x using -\t+2*\i] in {0,...,\steps}{      
				\thetastateflipped[-\x][0][1][#5]
			}
		}
	\end{scope}
}

\newcommandx{\eigenDL}[6][1=0,2=0,3=l,4=4,5=bertiniorange,6=topright]{
	\begin{scope}[shift={(#1,#2)}]
		\pgfmathsetmacro{\t}{#4}
		\ifthenelse{\equal{#3}{l}}{
			\eigenVL[0][0][l][\t][tr][init][#5][#6]
			\pgfmathsetmacro{\t}{#4-1}
			\rightriangle[1][0][\t][#5][#6]
			\drawinitstate[1][0][r][\t][#5]
		}{
			\begin{scope}[shift={(-0.5,0.5)}]
				\foreach \i[evaluate=\i as \x using \i, evaluate=\i as \y using \i] in {0,...,\t}{      
					\draw (\x,\y)--++(0.5,0);
					\draw[fill=white] (\x,\y) circle (0.15);
				}
			\end{scope}
		}
	\end{scope}
}

\newcommandx{\eigenDR}[6][1=0,2=0,3=l,4=4,5=bertiniorange,6=topright]{
	\begin{scope}[shift={(#1,#2)}]
		\pgfmathsetmacro{\t}{#4}
		\ifthenelse{\equal{#3}{r}}{
			\eigenVR[0][0][r][\t][tr][init][#5][#6]
			\pgfmathsetmacro{\t}{#4-1}
			\leftriangle[-1][0][\t][#5][#6]
			\drawinitstate[-1][0][l][\t][#5]
		}{
			\begin{scope}[shift={(0.5,0.5)}]
				\foreach \i[evaluate=\i as \x using \i, evaluate=\i as \y using \t-\i] in {0,...,\t}{      
					\draw (\x,\y)--++(0.5,0);
					\draw[fill=white] (\x+0.5,\y) circle (0.15);
				}
			\end{scope}
		}
	\end{scope}
}

\newcommandx{\idonpurity}[2][1=0,2=0]
{
	\begin{scope}[shift={(#1,#2)}]
		\draw[thick] (-0.5,0)--++(-0.1,0.1)--++(0,0.2)--++(0.1,-0.1);
		\draw[thick] (-0.5,0.4)--++(-0.1,0.1)--++(0,0.2)--++(0.1,-0.1);
		\draw[thick] (0.5,0)--++(0.1,0.1)--++(0,0.2)--++(-0.1,-0.1);
		\draw[thick] (0.5,0.4)--++(0.1,0.1)--++(0,0.2)--++(-0.1,-0.1);
	\end{scope}
}

\newcommandx{\swaponpurity}[2][1=0,2=0]
{
	\begin{scope}[shift={(#1,#2)}]
		\draw[thick] (-0.5,0)--++(-0.2,0.2)--++(0,0.6)--++(0.2,-0.2);
		\draw[thick] (-0.5,0.2)--++(-0.075,0.075)--++(0,0.2)--++(0.075,-0.075);
		\draw[thick] (+0.5,0)--++(+0.2,0.2)--++(0,0.6)--++(-0.2,-0.2);
		\draw[thick] (+0.5,0.2)--++(+0.075,0.075)--++(0,0.2)--++(-0.075,-0.075);
	\end{scope}
}

\newcommandx{\hook}[4][1=0,2=0,3=t,4=l]{
	\begin{scope}[shift={(#1,#2)}]
		\ifthenelse{\equal{#3}{t}}{
			\ifthenelse{\equal{#4}{l}}{\draw[thick] (0.5,-0.5) arc (45:-90:0.15);}{\draw[thick] (0.5,-0.5) arc (45:270:0.15);}
		}{\ifthenelse{\equal{#4}{l}}{\draw[ thick] (0.5,-0.5) arc (-45:90:0.15);}{\draw[ thick] (0.5,-0.5) arc (315:90:0.15);}
		}
	\end{scope}
}

\newcommandx{\hhook}[4][1=0,2=0,3=t,4=l]{
	\begin{scope}[shift={(#1,#2)}]
		\ifthenelse{\equal{#3}{t}}{
			\ifthenelse{\equal{#4}{l}}{\draw[thick] (0.5,-0.5) arc (-45:175:0.15);}{\draw[thick] (0.5,-0.5) arc (225:0:0.15);}
		}{\ifthenelse{\equal{#4}{l}}{\draw[ thick] (0.5,-0.5) arc (-45:180:-0.15);}{\draw[ thick] (0.5,-0.5) arc (45:-180:0.15);}
		}
	\end{scope}
}

\definecolor{FcolU}{rgb}{0.71,0.78,0.91}
\definecolor{colLines}{rgb}{0.31,0.31,0.31}
\definecolor{colVMPSLines}{rgb}{0.11,0.11,0.11}
\definecolor{IcolUc}{rgb}{0.71,0.41,0.42}
\definecolor{IcolU}{rgb}{0.71,0.8,0.76}
\definecolor{IcolVMPSc}{rgb}{0.73,0.69,0.7}
\definecolor{IcolVMPS}{rgb}{0.81,0.77,0.78}
\definecolor{colObs}{rgb}{1.,1.,1.}

\newcommandx{\eightlegs}[2][1=0,2=0]{
	\begin{scope}[shift={(#1,#2)}]
		\foreach \x in {1,...,8}{
			\draw (\x, 0)--++(0,0.25);
			\draw[fill] (\x,0) circle (0.05);
		}
		\foreach \x in {1,3}{
			\pgfmathsetmacro\result{2*\x-1} 
			\node () at (\result,-0.5) {$i_{\x}$};
			\pgfmathsetmacro\result{2*\x}
			\node () at (\result,-0.5) {$j_{\x}$};	
		}
		\foreach \x in {2,4}{
			\pgfmathsetmacro\result{2*\x} 
			\node () at (\result,-0.5) {$i_{\x}$};
			\pgfmathsetmacro\result{2*\x-1}
			\node () at (\result,-0.5) {$j_{\x}$};	
		}
	\end{scope}
}

\usepackage{xcolor}
\definecolor{blue50}{RGB}{127,127,255}
\definecolor{blue50}{RGB}{153,153,255}
\definecolor{blue10}{RGB}{230,230,255}
\definecolor{sblue50}{RGB}{163,193,173}
\definecolor{blue10}{RGB}{230,230,255}
\definecolor{brown50}{RGB}{210,148,148}
\definecolor{burgundy70}{RGB}{166,76,98}
\definecolor{orange50}{RGB}{255,210,127}
\definecolor{sblue50}{RGB}{163,193,173}

\usepackage{bbold}

\newcommand{\er}[1]{Eq.~\eqref{#1}}
\newcommand{\ers}[2]{Eqs.~(\ref{#1}-\ref{#2})}
\newcommand{\era}[2]{Eqs.~(\ref{#1}) and (\ref{#2})}
\newcommand{\eraa}[3]{Eqs.~(\ref{#1}), (\ref{#2}) and (\ref{#3})}

\newcommand{\Ers}[2]{Equations~(\ref{#1}-\ref{#2})}

\newcommand{\M}{\mathcal M}
\newcommand{\U}{\mathcal U}
\newcommand{\W}{\mathcal W}
\newcommand{\G}{\mathcal G}
\newcommand{\V}{\mathcal V}

\newcommand{\tM}{\widetilde\M}
\newcommand{\beq}{\begin{equation}}
\newcommand{\eeq}{\end{equation}}
\newcommand{\be}{\begin{equation}}
\newcommand{\ee}{\end{equation}}

\usepackage{mathtools}

\hypersetup{linkcolor=magenta,
    citecolor=blue}
    
\begin{document}

\title{Exact large deviations and emergent long-range correlations \\ in sequential quantum East circuits}

\author{Jimin Li}
\email{jl939@cam.ac.uk}
\affiliation{Department of Applied Mathematics and Theoretical Physics, University of Cambridge, Wilberforce Road, Cambridge CB3 0WA, United Kingdom}
\author{Bruno Bertini}
\affiliation{School of Physics and Astronomy, University of Birmingham, Birmingham B15 2TT, United Kingdom}
\author{Juan P. Garrahan }
\affiliation{School of Physics and Astronomy, University of Nottingham, Nottingham, NG7 2RD, UK}
\affiliation{Centre for the Mathematics and Theoretical Physics of Quantum
Non-Equilibrium Systems, University of Nottingham, Nottingham, NG7 2RD, UK}
\author{Robert L. Jack}
\affiliation{Department of Applied Mathematics and Theoretical Physics, University of Cambridge, Wilberforce Road, Cambridge CB3 0WA, United Kingdom}
\affiliation{Yusuf Hamied Department of Chemistry, University of Cambridge,
Lensfield Road, Cambridge CB2 1EW, United Kingdom}

\date{\today}

\begin{abstract}
Exploiting quantum measurements is a promising route for preparation of correlated quantum states.  We use methods from large deviation theory to solve this problem exactly for a specific system: the deterministic quantum East circuit with boundary measurements.  We show that conditioning on measurement outcomes generates a long-range correlated state, despite typical trajectories being trivial.  We derive the channel that optimally realizes the rare measurement trajectories, and establish a formal connection with the Petz recovery (time-reversal) map.  
We compute one- and two-point correlation functions in the conditioned state, revealing finite two-body correlations at arbitrarily large separations, and an underlying fractal structure, related to the Sierpiński triangle.
These results demonstrate explicitly how boundary measurements can be used to control bulk properties of a quantum system. 
\end{abstract}

\maketitle

\emph{Introduction} -- The current decade has witnessed the rise of quantum circuits---quantum many-body systems in discrete space and discrete time---as a new paradigm to understand out-of-equilibrium quantum matter~\cite{potter2022entanglement,fisher2023random,bertini2025exactly}. In broad terms, this has occurred for three main reasons: (i) in circuit systems one can discard much unnecessary detail while maintaining only essential physical ingredients, such as the locality of interactions; (ii) the symmetry between space and time realised by quantum circuits offers more avenues for theoretical analysis of their dynamics~\cite{potter2022entanglement,fisher2023random,bertini2025exactly,PhysRevLett.133.170402,PhysRevB.111.104315,yu2024hierarchical}; and (iii) circuits do model accurately certain actual quantum devices  ~\cite{evered2023high-fidelity,chen2024benchmarking,loschnauer2024scalable,fischer2024dynamical,acharya2025quantum,baccari2025average}.

\begin{figure}[t!]
    \centering
    \includegraphics[width=1\columnwidth]{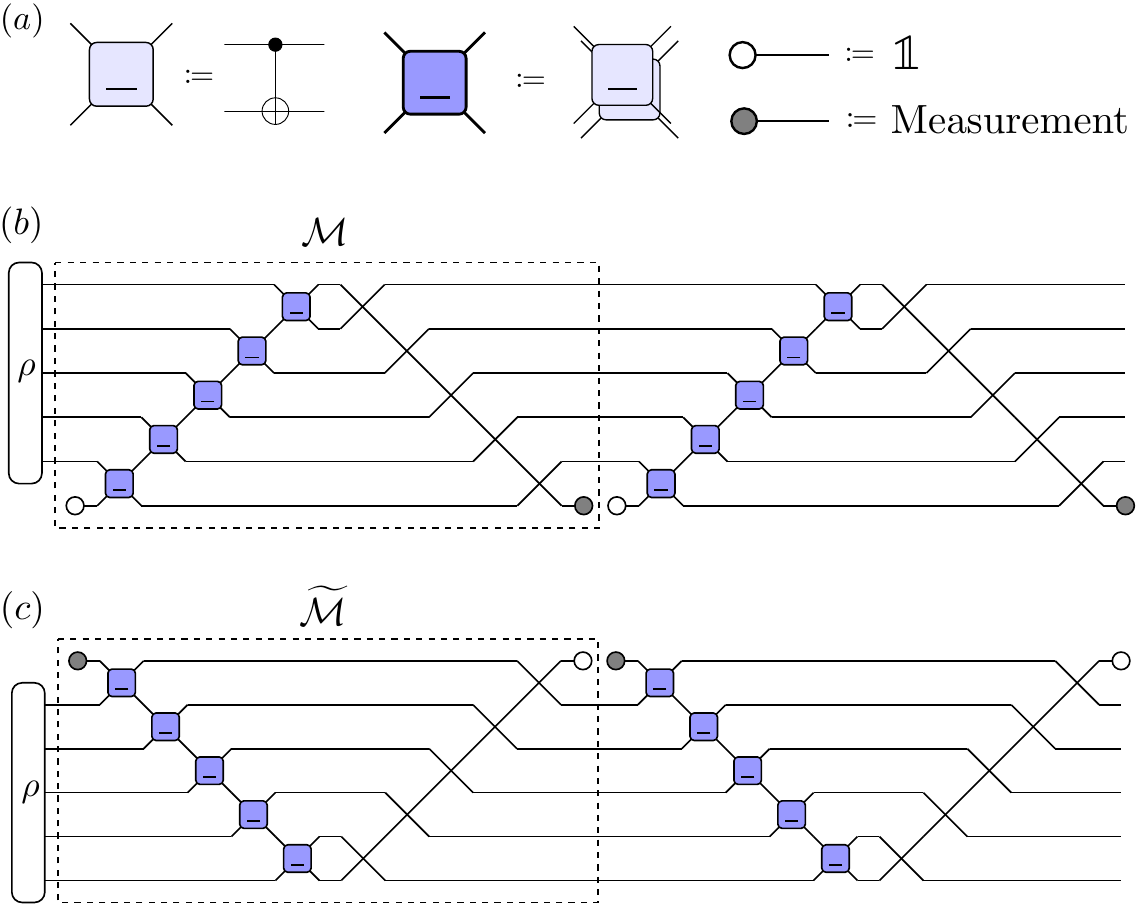}
    \caption{Sequential quantum East circuits. (a) CNOT gate, its folded counterpart, and identity and (diagonal) measurement matrices. (b) Sequential East circuit consisting of two iterations of the channel $\M$.  The circuit acts on $L+1$ sites ($L=5$ in the figure), which couple an input on the first $L$ sites to a maximally mixed ancilla. Measurements are performed at the ancilla. (c) Alternative circuit in which the ancilla is prepared in a non-trivial mixed state and traced after each application of the channel $\tM$.}
    \label{fig:fig1}
\end{figure}

A particularly significant research direction facilitated by the discrete spacetime setting, has been the interplay of unitary dynamics and measurements. This raises both conceptual and practical questions for quantum dynamics and quantum computation: measurements can tame the growth of entanglement induced by unitary evolution, leading to much-studied phase transitions in conditioned states, see e.g.\ Refs.~\cite{skinner2019measurement-induced,li2019measurement-driven,hoke_measurement-induced_2023}; 
combining unitary dynamics and measurements can also allow for the preparation of highly entangled states at low circuit depths, see e.g.\ Refs.~\cite{piroli2021quantum,zhu2023nishimoris,PhysRevLett.133.230401, iqbal2024topological, gopalakrishnan2023push-down,lu2025spacetime,lloyd2025quantum,PhysRevResearch.6.033147} and references therein. 

A convenient setting for such investigations is based on so-called ``collision models'' \cite{ciccarello2022quantum, cilluffo2021microscopic, cech2023thermodynamics, 
cech2025space-time,cech2025revealing,RevModPhys.93.045003,schon2005sequential,schon2007sequential,banuls2008sequentially,wei2022sequential} 
(a discrete time version of the input-output formalism \cite{belavkin1990a-stochastic,carmichael1993an-open,gardiner2004quantum}).  Here, a quantum system is made to interact with a different ancilla at every step of the evolution, after which the ancilla is measured. Like other versions of quantum Markov chains, a collision model defines a quantum channel for the dynamics of the average state; it also yields an {\em unravelling} into quantum trajectories for the stochastic evolution of the state conditioned on the measurements~\cite{plenio1998the-quantum-jump,breuer2002the-theory,gardiner2004quantum}. Specifically, given the channel that defines the stochastic dynamics, the statistics of measurements and conditioned states are encoded by the {\em tilted channel}~\cite{garrahan2010thermodynamics,esposito2009nonequilibrium,garrahan2018aspects}. This works as follows: the outcomes of repeated measurements are summed over time, for each trajectory its probability is exponentially reweighted by these totals, and these reweighted probabilities added up.  Hence one obtains the moment generating function of the counts, which coincides with the trace of the exponential of the tilted generator. Furthermore, the {\em tilted ensemble} of trajectories captures rare events of the dynamics via a form of post-selection, leading to interesting properties in the conditioned state.  In principle, this post-selected ensemble can be sampled efficiently through a quantum Doob transformation \cite{garrahan2010thermodynamics,carollo2018making}, i.e., the proper stochastic dynamics that generates the exponentially tilted probabilities. Obtaining this {\em Doob dynamics} is generally difficult as it requires knowledge of the spectrum of the tilted generator, and as far as we are aware, there is no example of an exact solution to this problem in a truly many-body quantum setting, preventing one to study any associated measurement-induced long-range correlations.  

Here, we consider a particular many-body collision model in which the analysis of the tilted channel can be carried out exactly and explicitly. The system consists of $L$ system qubits and one ancilla qubit per time step. At each step the qubits are evolved by a sequential circuit of CNOT (i.e.\ the gates of the quantum East circuit at its deterministic point ~\cite{berenstein2021exotic,klobas2024exact,bertini2024localized,bertini2024exact}) and SWAP gates followed by a projective measurement of the ancilla in the computational basis, see Fig.~\ref{fig:fig1}. Note that in our setup the unitary evolution is implemented by an automaton circuit ~\cite{PhysRevB.100.214301, PRXQuantum.2.010329, Han2023entanglement, feng2025dynamics, PhysRevX.15.011015, szaszschagrin2025entanglementdynamicspagecurves, mcdonough2025bridgingclassicalquantuminformation, bertini2025randompermutationcircuitsquantum}. On average, the system is evolved by a unital channel, ${\mathcal M}$, whose trivial steady state lacks any correlations. However, as we prove exactly below, the behaviour at the level of fluctuations is far from trivial: exponentially tilting according to measurement outcomes yields a conditional state on the system that is long-range correlated, and described by a linear-depth circuit. Furthermore, via the exact quantum Doob transform, we find a new quantum circuit that reproduces these correlations in its typical trajectories and therefore in its steady state. Besides providing a controlled many-body setting where to study exactly the interplay of measurements and unitary dynamics, our results also provide a realisable system with which to benchmark current quantum devices. 

\emph{Setup} -- We consider a quantum channel $\M$ defined as the spacetime circuit shown in Fig.~\ref{fig:fig1}(a,b). Throughout we use the standard diagrammatic techniques of quantum circuits, and in particular we work in the \emph{folded representation}, see e.g.\ Ref.~\cite{bertini2025exactly}, so that shapes with legs protruding from two (left and right) sides indicate superoperators and those with legs on one only (left or right) side are matrices (or operators). Figure~\ref{fig:fig1} shows a circuit of $L=5$ qubits (the general $L$ setup should be evident) but all our results are for arbitrary $L$.  

The circuit of Fig.~\ref{fig:fig1}(a,b) describes the following dynamics: (i) prepare an ancilla in the maximally mixed state $\rho_{\rm anc}={\mathbb 1}$; (ii) evolve the system and ancilla together using $L$ CNOT gates and $L$ SWAP gates applied as in the figure; (iii) measure the ancilla in the computational basis $\{\ket{0}, \ket{1}\}$. The sequence (i-iii) defines one ``time step'' where time $t$ advances by $1$. We write $Q_k(t)$ for the number of measurement outcomes of type $k\in\{0,1\}$ up to time $t$, noting that $Q_0(t)+Q_1(t)=t$. 

To post-select states conditioned on $Q_k(t)$, we employ large-deviation methods for quantum open dynamics \cite{garrahan2010thermodynamics,carollo2018making,carollo2019unraveling,li2025efficient}. Specifically, in order to post-select to change the average of $Q_0(t)-Q_1(t)$ we tilt the channel by a {\em counting field} $s$, such that the probabilities of measurement outcomes are reweighted by ${\rm e}^{s[Q_0(t)-Q_1(t)]}$. In this way, $s>0$ favours $Q_0(t)-Q_1(t)$ larger than typical, while $s<0$ disfavours it.  In the notation of Fig.~\ref{fig:fig1}(b), this corresponds to replacing the grey bullet in the figure by a {\em measurement (or tilting) matrix}
\beq
\fineq{
\draw[ thick] (0,0) -- (1, 0);
\draw[fill=bertiniblue,draw=black, thick]  (1,0) circle  (0.15); }
\; = 
\begin{pmatrix} e^{s} & 0 \\ 0 & e^{-s} \end{pmatrix} 
=: m_s. 
\label{eq:ms}
\eeq
The resulting tilted channel is then
\beq
\M_s =  \fineq[-0.8ex][1][1]{
\begin{scope}[rotate=-45,scale=0.75]
\tsfmatV[-5][4][r][1][][][blue50][topleft]
\tsfmatV[-4][5][r][1][][][blue50][topleft]
\tsfmatV[-3][6][r][1][][][blue50][topleft]
\tsfmatV[-2][7][r][1][][][blue50][topleft]
\tsfmatV[-1][8][r][1][][][blue50][topleft]
\draw[fill=white, draw=black, thick]  (-1.5 + 1 , 8.5 +1) circle  (0.15);
\draw[fill=bertiniblue, draw=black, thick]  (-5.5 ,4.5) circle  (0.15);
\end{scope}
} \; .
\label{eq:Ms}
\eeq

\emph{Dominant eigenmatrices of $\M_s$} -- For large $t$, the behaviour of post-selected trajectories is controlled by the leading eigenmatrices of the tilted channel and of its adjoint, $\M_s^\dag$. Using the properties of CNOT gates, we show~\cite{see-supplemental} that these leading eigenmatrices are
\begin{equation}
\label{eq:M-fixed}
\M_{{s}}[\mathbb{1}] 
  = e^{\theta({s})} \mathbb{1}
\qquad
\M_{{s}}^{\dagger}\big[\rho^{\triangleleft}_{s}\big] 
  = e^{\theta({s})} \rho^{\triangleleft}_{s}, 
\end{equation}
with  
\begin{equation}
  \label{eq:theta}
  \theta({s})=\ln{(2 \cosh{s})} = \ln \Tr [m_s] , 
\end{equation}
and where
\begin{equation}
\rho^{\triangleleft}_{s} := \U^{\triangleleft}\big( m_s^{\otimes L} \big) e^{-L\theta(s)}, 
\label{eq:psi-triag}
\end{equation}
is defined from the action of the super-operator
\begin{align} \label{eq:U-triag}
\U^{\triangleleft} =
\text{
\fineq[-0.8ex][1][1]{
\begin{scope}[scale=0.55]
\begin{scope}[rotate= -135 , shift={( -5.7 , -6 )}]
\fill[fill=blue!10, draw=black, dashed ]
  (6.4,6.4) -- (6.4,-0.4) -- (-0.4,-0.4) -- cycle;
\end{scope}
\tsfmatV[-4][3][r][1][][][blue50][topright]
\tsfmatV[-3][2][r][1][][][blue50][topright]
\tsfmatV[-2][1][r][1][][][blue50][topright]
\tsfmatV[-1][0][r][1][][][blue50][topright]
\tsfmatV[-2][3][r][1][][][blue50][topright]
\tsfmatV[-1][2][r][1][][][blue50][topright]
\tsfmatV[-2][5][r][1][][][blue50][topright]
\tsfmatV[-3][4][r][1][][][blue50][topright]
\tsfmatV[-1][6][r][1][][][blue50][topright]
\tsfmatV[-1][4][r][1][][][blue50][topright]
\draw[thick] (-0.5,1.5) -- (-0.5,2.5);
\draw[thick] (-0.5,3.5) -- (-0.5,4.5);
\draw[thick] (-0.5,5.5) -- (-0.5,6.5); 
\draw[thick] (-4.5, 0.5 ) -- (-4.5, 3.5 );
\draw[thick] (-3.5, 0.5 ) -- (-3.5, 2.5 );
\draw[thick] (-2.5, 0.5 ) -- (-2.5, 1.5 );
\draw[thick] (-4.5, 4.5 ) -- (-4.5, 7.5 );
\draw[thick] (-3.5, 5.5 ) -- (-3.5, 7.5 );
\draw[thick] (-2.5, 6.5 ) -- (-2.5, 7.5 );
\draw[thick] (-0.5,0.0) -- (-0.5,0.5);
\draw[thick] (-1.5,0.0) -- (-1.5,0.5);
\draw[thick] (-2.5,0.0) -- (-2.5,0.5);
\draw[thick] (-3.5,0.0) -- (-3.5,0.5);
\draw[thick] (-4.5,0.0) -- (-4.5,0.5);
\draw[thick] (-0.5, 7.5) -- (-0.5, 8);
\draw[thick] (-1.5, 7.5) -- (-1.5, 8);
\draw[thick] (-2.5, 7.5) -- (-2.5, 8);
\draw[thick] (-3.5, 7.5) -- (-3.5, 8);
\draw[thick] (-4.5, 7.5) -- (-4.5, 8);
\end{scope}
}
}.
\end{align} 

From \ers{eq:M-fixed}{eq:psi-triag} we deduce that $\theta(s)$ is the scaled cumulant generating function for the difference in $0$ and $1$ measurements at long times $t$. Its simple form arises because these measurements are independent Bernoulli random variables in the steady state of the original channel $\M$.  Despite this simplicity, we now show that the states conditioned on these measurements are rich and complex.

\begin{figure}[t!]
\centering
\includegraphics[width=1\columnwidth]{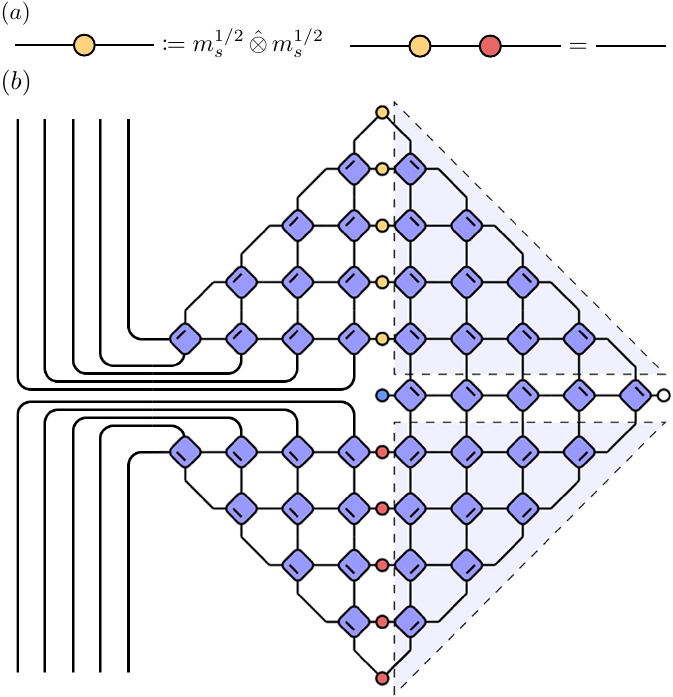}
\caption{Doob channel. (a) Measurement super-operators Eq. \eqref{eq:mm} and their inverses. (b) Tensor network representation of the Doob channel $\M^{\rm D}_s$.
}
\label{fig:fig2}
\end{figure}

\emph{Quantum Doob transform} --
The tilted channel is not trace-preserving, $\M_{{s \neq 0}}^\dag[\mathbb{1}] \neq \mathbb{1}$.  We therefore apply the quantum Doob transformation \cite{garrahan2010thermodynamics,carollo2018making} to obtain a physical (trace-preserving) channel $\M^{\text{D}}_{{s}}$ that reproduces the tilted ensemble of trajectories. The standard procedure leads to~\cite{see-supplemental}
\beq
\M^{\text{D}}_{{s}}  
  := e^{-\theta(s)} \, \V_s \circ \M_{s} \circ \V_s^{-1}, 
\label{eq:M-doob}
 \eeq
 where $\circ$ denotes composition of superoperators, and 
 \beq
 \begin{aligned}
  \V_s(\cdot) 
    & = 
    (\rho^{\triangleleft}_s)^{1/2} (\cdot) (\rho^{\triangleleft}_s)^{1/2} 
    \\
    &=  
    e^{-L\theta(s)} \U^\triangleleft \circ (m_s^{1/2} {\hat \otimes} \, m_s^{1/2} )^{\otimes L} \circ \U^\triangleleft(\cdot)
    \,.
\end{aligned}
\label{eq:calVs}
\eeq
Note that $m_s^{1/2}$ is well defined ($m_s$ is positive definite), and $\, {\hat \otimes} \,$ denotes the ``fold'' tensor product, meaning that action of the single-qubit superoperator is
\begin{equation}
  \label{eq:mm}
(m_s^{1/2}{\hat \otimes} \, m_s^{1/2} )[\cdot] = m_s^{1/2} (\cdot) \, m_s^{1/2} .  
\end{equation}

Fig.~\ref{fig:fig2} shows \er{eq:M-doob} in tensor network form, with $\V_s(\cdot)$ given by \er{eq:calVs}. The diagram involves four triangles, each representing $\U^\triangleleft$ [cf.\ \er{eq:U-triag}], and $2L$ two-legged circles, each representing $m_s^{1/2} {\hat \otimes} \,  m_s^{1/2}$, since each $\V_s$ in Eq. \eqref{eq:M-doob} involves two of the former and $L$ of the latter.  

The Doob channel is trace-preserving: its right fixed point (steady state) is $\rho^{\triangleleft}_{s}$ and that of its adjoint is the identity. \Ers{eq:psi-triag}{eq:U-triag} express the steady state $\rho^{\triangleleft}_{s}$ as a linear-depth circuit which encodes long-ranged system correlations, despite the fact that the original (untilted) channel was unital and the measurement outcomes have simple statistics. We will characterise these correlations explicitly below, but we first make a useful connection to a form of time reversal. 

\emph{Doob transform as time-reversal} -- Consider an alternative dynamics given by the linear sequential circuit shown in Fig.~\ref{fig:fig1}(c). In this circuit the ancilla is prepared in mixed state $\rho_{\rm anc}={\rm e}^{-\theta(s)} m_s$ [cf.\ the identity for the original circuit of Fig.~\ref{fig:fig1}(b)], and the sequence of CNOT gates runs from top to bottom. The circuit corresponds to iterated application of the new channel
\beq
\label{eq:tMs}
\tM_s
  = e^{-\theta(s)} \; \fineq[-0.8ex][1][1]{
\begin{scope}[rotate=-45, scale=0.75]
\tsfmatV[-5][4][r][1][][][blue50][topright]
\tsfmatV[-4][5][r][1][][][blue50][topright]
\tsfmatV[-3][6][r][1][][][blue50][topright]
\tsfmatV[-2][7][r][1][][][blue50][topright]
\tsfmatV[-1][8][r][1][][][blue50][topright]
\draw[fill=white,draw=black, thick]  (-1.5 + 1,8.5+1) circle  (0.15);
\draw[fill=bertiniblue,draw=black, thick]  (-5.5,4.5) circle  (0.15);
\end{scope}
}. 
\eeq
This channel is parameterised by the field $s$ but, contrary to $\M_s$, it is completely positive and trace-preserving. This is seen by comparing \er{eq:tMs} with \er{eq:Ms}, which shows that 
\begin{equation}
\label{eq:tMsMsd}
  \tM_s = e^{-\theta(s)} \M^\dag_s , 
\end{equation}
so that 
\begin{equation}
  \label{eq:mtilde}
  \tM_s[\rho^\triangleleft_s] = \rho^\triangleleft_s
  \qquad
  \tM_s^{\dagger}\big[\mathbb{1}] 
  = \mathbb{1} .
\end{equation}
The non-trivial mixed state of the ancilla enables these non-trivial correlations in the {\em typical} dynamics of $\tM_s$. 

Comparing \era{eq:tMsMsd}{eq:M-fixed} suggests that the typical dynamics of the new channel and the rare events of the original channel are related by time-reversal. This is proved as follows. For a generic channel $\G$ with unique fixed point $\rho_\infty$, the associated time-reversed (stationary) dynamics is given by the {\em Petz recovery map} \cite{petz1986sufficient,kwon2022reversing,bai2025quantum}
\beq
\label{eq:Petz}
\G^{\rm P}[\cdot] := 
  \rho_\infty^{1/2} \, \G^\dag[\rho_\infty^{-1/2} (\cdot) \rho_\infty^{-1/2}]  \, \rho_\infty^{1/2} .
\eeq
For $\G=\tM_s$ and using \eraa{eq:M-doob}{eq:tMsMsd}{eq:mtilde} we get 
\beq
  \label{eq:MDP}
 \M^{\text{D}}_{{s}} = \widetilde{\M}^{\text{P}}_s. 
\eeq 
Physically, this means that quantum trajectories of the conditional state under the Doob channel $\M_s^{\rm D}$ are those of $\widetilde{\M}_s$ but time-reversed, and their stationary states coincide. 

More generally, the connection to the Petz map highlights the complexity of the post-selection problem for $\M_s$: it can be solved by the Doob channel of Fig.~\ref{fig:fig2} (where the difficulty is finding the leading left eigenmatrix of $\M_s$), or by building the simpler channel $\tM_s$ and then time-reversing its dynamics (where the difficulty is computing the Petz map). Either procedure leads to the same $\rho_s^\triangleleft$ and the same tilted ensemble of conditional state trajectories.

\emph{Time-inhomogeneous tilting} -- We can show \cite{see-supplemental} that the steady state in Eq.~\eqref{eq:mtilde} is reached in a \emph{finite} number of applications of $\tM_s$ (or alternatively $\M^{\text{D}}_{{s}}$), specifically in ${t=L}$ time steps. This implies that, when thought of as a matrix, $\tM_s$ has a special structure: its spectrum is $\{0,1\}$, the eigenvalue $1$ is non-degenerate, and the largest Jordan block corresponding to the eigenvalue 0 has size $L$. In fact an even stronger fact holds \cite{see-supplemental}: as long as we use tilting matrices of the form Eq. \eqref{eq:ms} the same structure is found for all $s$. This means that even when considering different tilt parameters $(s_1 \cdots s_t)$ at each time step $t$, the state at time $t\geq L$ is given by a from similar to that of \er{eq:psi-triag}, 
\begin{align}
  \label{eq:rtL}
\rho_{t \geq L} 
  &=
  \tM_{s_t} [\cdots \tM_{s_1}[ \rho_0 ]] 
  \\ 
  &=
  \U^{\triangleleft}( m_{s_t} \otimes  \cdots \otimes m_{s_{t-L+1}})
  e^{- \sum_{t'=t-L+1}^t \theta(s_{t'})}  
  , 
  \nonumber
\end{align}
which is independent of the initial condition $\rho_0$. This means that after $L$ steps this process loses all memory of the initial condition and only retains information of the last $L$ tilt matrices.

\begin{figure}[t!]
    \centering
    \includegraphics[width=1\columnwidth]{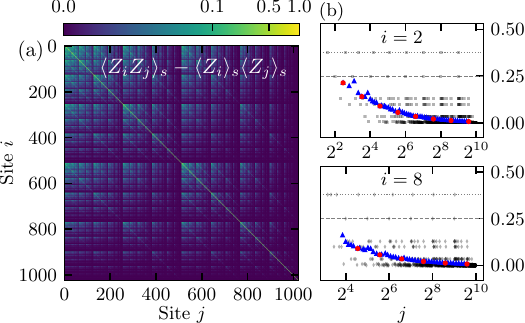}
    \caption{
      (a)~Correlation function $C_s(i,j)$ for $s=0.9$ and $L = 1024$. 
      (b)~$C_s(i,j)$ as a function of $j > i$ for fixed  $i=2$ (top) and $i=8$ (bottom). The gray dashed and dotted horizontal lines indicate the values of $C_s(2^m,2^n)$ and $C_s(2^m,2^n + 2^m)$. The red circles are the averages $[C_s(i,\cdot)]_{2^n}^{2^{n+1}}$. The blue triangles are $[C_s(i,\cdot)]_{2^{(n/8)}}^{2^{(n/8)+1}}$.
    } 
    \label{fig:fig3}
\end{figure}

\emph{Long-ranged correlations in the conditioned state} -- 
We now show that in contrast to the trivial stationary state of $\M$, when conditioning on measurements, the state develops long range correlations. Consider a system operator $O$. Its expectation value at some intermediate time $t$ over post-selected trajectories that run between times $0$ and $T$ can be obtained from the tilted channel as \cite{garrahan2010thermodynamics,carollo2018making}
\begin{equation}
  \langle O(t) \rangle_{\rm s} := \frac{ \Tr\left( \M_{ s}^{T-t}\! \left[ O \M_{s}^{t}[\rho_0]\right] \right) }{ \Tr( \M_{ s}^{T}[\rho_0]) } .
\label{eq:O-tilt}
\end{equation} 
In the limit of long times ($t, T \to \infty$) we can use the fixed point structure, \ers{eq:M-fixed}{eq:psi-triag}, to obtain at stationarity
\begin{equation}
  \label{eq:Oav}
\langle O \rangle_{\rm s}  
  = 
  \Tr\left[\U^\triangleleft(O) \, m_s^{\otimes L}\right] {\rm e}^{-L\theta(s)} .
\end{equation}  
Given the finite convergence time of the circuit, cf.\ \er{eq:rtL}, the above expression actually holds for any {$L \leq t \leq T-L$}. Also, due to \er{eq:MDP}, these are also the steady state correlations of the circuit $\tM_s$.

Consider the operator $Z_i$ that acts as Pauli $\sigma^z$ on site $i$ and trivially on all other sites. One can show \cite{see-supplemental} (where we label sites $i = 0, \ldots, L-1$)
\begin{equation}
  \label{eq:ztilde}
  \tilde{Z}_i := \U^\triangleleft(Z_i) = \prod_{k=0}^i Z_k^{ \beta_{i,k} } ,
\end{equation}
where $\beta_{i,k}=\binom{i}{k} \text{ mod } 2$ is the Sierpiński triangle. With \er{eq:ztilde}, using \er{eq:Oav} we can calculate the one- and two-point correlations of $Z_i$, 
\begin{align}
  \label{eq:Z}
  \!\!\langle Z_i \rangle_{{s}} & = \exp \left[ q(s) \sum_{k=0}^i \beta_{i,k} \right] ,
  \\
  \label{eq:ZZ}
  \!\!\langle Z_i Z_j \rangle_{{s}} & =  \exp\left\{ q(s) \!\!\left[ \sum_{k=0}^i (\beta_{i,k} \oplus \beta_{j,k} ) + \!\!\!\sum_{k=i+1}^j \!\!\!\beta_{j,k}\! \right] \right\} ,
\end{align}
where we consider $i<j$, $s>0$, we have defined $\beta \oplus \beta'=(\beta+\beta') \text{ mod } 2$, and 
$q(s)=\ln \tanh(s)$. For $s<0$ the same formulae hold but $q$ is complex [${\rm Im}(q)=\pi$]. Therefore, while the expectation values are still real they can be negative. For $s\to0$, ${\rm Re}(q)\to-\infty$ and the expectation values above vanish (except for $i=j$), reflecting that $\rho^\triangleleft_s \to \mathbb{1}$ in this limit. In the limit $s \to \infty$ we have $q(s) \to 0$ and both expectation values approach one reflecting that $\rho^\triangleleft_s \to |0\rangle\!\langle0|^{\otimes L}$.

The result for $\langle Z_i \rangle_{{s}}$ means that the system has no ``bulk'', i.e., far from the origin, this one-point function is close to zero at most sites, but there are some special sites where it stays of order unity: for example, $\langle Z_i\rangle_s={\rm e}^{2q(s)}$ for $i=2^m$. 

To understand the ``typical'' behaviour of the correlators, we average them over spatial windows, as follows.  
Recalling Eq.~\eqref{eq:Z} we consider $V_j= \sum_{k=0}^j \beta_{j,k}$ which is the number of $1$'s in the $j$th row of the Sierpiński's triangle. 
 We take $L=2^m$ and average the $k$th power of $V_j$ over rows $j$ between $0$ and $L-1$ to obtain~\cite{see-supplemental}
 \beq
[V^k]_0^{L}  
  = 
  \left(\frac{1+2^k}{2}\right)^m = L^{d_{\frak f}(k)-1}
\eeq
where  $[(\cdot)]_a^b := (b-a)^{-1} \sum_{j=a}^{b-1} (\cdot)_j$ is the windowed average,
and $d_{\frak f}(k) := \ln(1+2^k)/\ln 2$ [$d_{\frak f}(1) \approx 1.584$ is the fractal dimension of the Sierpiński triangle]. This gives the following averaged one-point function 
\beq
[\expval{Z}_s]_0^{L} 
  = 
  \frac{1}{2^m}\sum_{j=0}^m \binom{m}{j} e^{2^j q(s)} \geq \frac{e^{q(s)}}{L}. 
\eeq

A similar fractal structure is also observed in the connected two-point function 
\beq
C_s(i,j) = \langle  Z_i  Z_j \rangle_s -\langle  Z_i \rangle_s\langle Z_j \rangle_s.
\eeq
For instance, Fig.~\ref{fig:fig3} shows that there exist pairs of sites with arbitrarily large separation with $C_s$ of the order unity, indicating infinite-ranged correlations. From Eq.~\eqref{eq:ZZ} we have that for positive integers $m\neq n$ 
\begin{align}
\label{eq:twopointresult}
C_s(2^m,2^n)  & = e^{2q(s)} - e^{4q(s)}
  \\
\label{eq:twopoint2}
C_s(2^m,2^n+2^m) & = e^{2q(s)} - e^{6q(s)}
\end{align}
which do not decay as $m,n$ increase, see Fig~\ref{fig:fig3}.

On the other hand, taking $i=2^m$ and averaging $j$ in the window $[2^n, 2^{n+1}-1]$ with $n>m$ we find \cite{see-supplemental}
\begin{align}
  [C_s(2^m,\cdot)]_{2^n}^{2^{n+1}} 
  &=
  \frac{1}{2^n}\sum_{j=0}^{n-1} 
  \binom{n-1}{j} 
  \left( e^{2^{j+2} q(s)}
  \right.
  \\
  &
  \left. 
  - 2\sinh[2q(s)] e^{2^{j+1} q(s)} \right)
  + 
  \frac{e^{2(2^{n}+1)q(s)} }{2^n} .
  \nonumber
\end{align}
This average provides an estimate of the typical correlations in the conditioned state of  Eq. \eqref{eq:psi-triag}, see Fig.~\ref{fig:fig3}. These spatially averaged correlations
decay algebraically with the window size
(which is $2^n)$; they are independent of $m$ so this two-body correlation is primarily determined by the absolute distance from the origin and not the distance between the qubits.
We note that while $\rho^\triangleleft_s$ has long-ranged correlations, these are classical in nature as the state is separable (it is diagonal in the computational basis).

\emph{Discussion} -- Here we have calculated exactly the dynamical large deviations of measurements at the boundary of a sequential quantum East circuit. Despite the fact that the stationary state is trivial and the measurements a Bernoulli process, we have shown that when the final state is conditioned (specifically, exponentially tilted) on the measurement outcomes it contains long-range correlations. This is a form of boundary-bulk correspondence, where measurements on the boundary control the structure on the bulk of the system.
Our result here is one of the very few examples of an exact characterisation of dynamical large deviations in a genuine many-body system. 

We have also found an interesting connection to time-reversal, where the ``Doob dynamics'' that optimally samples the rare events of the circuit corresponds to the Petz recovery map that time-reverses the dynamics of a different circuit whose environment is structured. As for the large deviations, this result is a rare example of the exact solution of a Petz map in a many-body problem. 

The circuit we study can be realised in current quantum devices as it requires only CNOT and SWAP gates together with mid-circuit measurements \cite{rudinger2021characterizing}. Our results provide a way to create long-range correlated states simply by controlling a boundary ancilla. Being exact, our results can also serve as a benchmark for experimental devices. While the correlations we have shown here are only classical, a similar setting could be used to generate genuine entanglement by simply rotating the measurement basis. We hope to report on this and other extensions in future work.

\begin{acknowledgments}
JL acknowledges support by the UKRI Grant No.\ EP/Z003342/1. BB acknowledges financial support from the Royal Society through the University Research Fellowship No.\ 201101. JPG acknowledges financial support from EPSRC Grant No.\ EP/V031201/1 and Leverhulme Trust Grant No.\ RPG-2024-112. 
\end{acknowledgments}

\bibliography{bibliography-08092025.bib, extra.bib}

\onecolumngrid
\break
\begin{center}
  {\bf\large Supplemental Material: \\ Exact large deviations and emergent long-range correlations in sequential quantum East circuits}
\end{center}
\onecolumngrid

\renewcommand{\theequation}{S\arabic{equation}}
\renewcommand{\thefigure}{S\arabic{figure}}

\newcounter{equationSM}
\newcounter{figureSM}
\newcounter{tableSM}
\stepcounter{equationSM}
\setcounter{equation}{0}
\setcounter{figure}{0}
\setcounter{table}{0}
\setcounter{section}{0}
\makeatletter
\renewcommand{\theequation}{\textsc{sm}-\arabic{equation}}
\renewcommand{\thefigure}{\textsc{sm}-\arabic{figure}}
\renewcommand{\thetable}{\textsc{sm}-\arabic{table}}

In this supplemental material, we provide the following to complement the main text: 
\begin{itemize}
    \item In Sec. \ref{sec:setup}, we provide the detailed definitions of the circuits of interest, and the standard diagrammatic notions. 
    \item In Sec. \ref{sec:largde_deviation}, we review the applications of Level-$1$ large deviation methods to quantum operations.
    \item In Sec. \ref{sec:fixed_points}, we discuss the tilted quantum channel and show the exact dominant eigenmatrices.
    \item In Sec. \ref{sec:Doob}, we discuss the quantum Doob transformation. 
    \item In Sec. \ref{sec:Inhomogeneous}, we provide of the details of time-inhomogeneous tilting setup.
    \item In Sec. \ref{sec:Observables}, we provide the calculations of observables in the conditioned state. 
\end{itemize}

\section{Quantum East Circuits}
\label{sec:setup}
In this section, we introduce the quantum East circuits studied in the main text, which consist of sequential quantum circuits with boundary measurement, and summarise the standard graphical notations for quantum circuits. 

\subsection{Sequential quantum circuits with boundary measurement}

Throughout this work, we focus on (1+1)-dimensional quantum circuits that consist of qubits with a local Hilbert space of $\mathcal{H}_{i} = \mathbb{C}^{2}$, where the subscript $i$ labels the site index. The building block of circuits we are considering is the CNOT gate, 
\begin{equation}
    U^{\text{CNOT}} = P^0 \otimes \mathbf{1} + P^1 \otimes X,
\label{eq:CNOT_gate}    
\end{equation}
where $\{ X, Y , Z \}$ are the spin-$\frac{1}{2}$ Pauli matrices, and $P^{0} = \left( \mathbf{1} + Z \right)/2$ and $ P^{1} = \left( \mathbf{1} - Z \right)/2$ are the projection operators in the computational basis. To make a connection with previous work on the quantum East circuits, the gate considered in Ref \cite{bertini2024exact} is $\left( X \otimes X \right) U^{\text{SWAP}}U^{\text{CNOT}}U^{\text{SWAP}}  \left( X \otimes X \right)
=  \mathbf{1}\otimes P^1 +  X \otimes P^0$, where $U^{\text{SWAP}}$ is the SWAP gate. 

A \textit{sequential quantum circuit (SQC)} is a special kind of linear-depth quantum circuit, where every layer only consists of non-identity gates on a sub-region of the domain \cite{schon2005sequential,schon2007sequential,banuls2008sequentially,wei2022sequential}. Each layer of the SQC considered in this work consists of either a single CNOT gate $ U^{\text{CNOT}}_{i,i+1}$ acting on the $i$ and $i+1$ sites, or a similar SWAP gate $ U^{\text{SWAP}}_{i,i+1}$.  

The circuit is defined by applying a sequence of CNOT gates as
\begin{equation}
    U_{\text{SQC}} \coloneq U^{\text{CNOT}}_{0,1} \, U^{\text{CNOT}}_{1,2} \cdots U^{\text{CNOT}}_{L-1,L} 
    \label{eq:SQC}
 \end{equation}
 we also define an operator
 \begin{equation}
    W_{\text{SQC}} \coloneq U^{\text{SWAP}}_{L-1,L} \, U^{\text{SWAP}}_{L-2,L-1} \cdots U^{\text{SWAP}}_{0,1}
\end{equation}
which realises the \textit{shift} operation on $L+1$ sites \cite{piroli2020quantum}.

The system considered in Fig.~\ref{fig:fig1}(b) has $L$ qubits and describes the following dynamics. 
Starting with a system density matrix $\rho \in \text{End}(\mathcal{H}^{L} \otimes \mathcal{H}^{L} )$, we first couple the system to an ancilla qubit in the maximally mixed state; second, apply the sequential circuit; third, measure the ancilla in the computational basis.  
The (un-normalised) conditional state of the system associated with measurement outcome $k=0,1$ is then
\beq
\Tr_L \biggl[  {P}^{(k)}_{L} \, \W_{\text{SQC}} \circ \U_{\text{SQC}} (\rho \otimes \mathbb{1})  \biggr] ,
\label{eq:cond-state-left}
\eeq
{where ${P}^{(k)}_{L}=\mathbb{1}\otimes |k\rangle\langle k|$ is a projector for state $|k\rangle$ of the ancilla (which is the $L$th qubit)}; the trace is over the ancilla, and we introduced superoperators
\beq
\U_{\text{SQC}} (\cdot) = U_{\text{SQC}}(\cdot) U_{\text{SQC}}^\dag, \qquad
\W_{\text{SQC}} (\cdot) = W_{\text{SQC}}(\cdot) W_{\text{SQC}}^\dag .
\label{eq:definesuperoperator}
\eeq
To define the channel ${\cal M}$ we average over measurement outcomes to obtain 

\beq
\M [\rho] \coloneq  \Tr_{L} \biggl[ \W_{\text{SQC}} \circ \U_{\text{SQC}} (\rho \otimes \mathbb{1})  \biggr],
\eeq

This quantum channel $\M$ is a \textit{completely positive and trace preserving} (CPTP) linear map~\cite{nielsen2010quantum}, note that $\M^{\dagger}[\mathbb{1}] = \mathbb{1}$. Additionally, we observe that $\mathcal{M}[\mathbb{1}] = \mathbb{1}$, which makes the quantum channel \textit{unital}. Unital quantum channels are quantum analogues of bistochastic matrices in classical Markov matrices. 

\subsection{Complementary channel}
In addition to the SQC with boundary measurement setup, we also introduce the
 complementary channel of Fig. \ref{fig:fig1}(b), which plays an important role when considering the quantum Doob transformation. 

We define
\begin{equation}
   \widetilde{U}_{\text{SQC}} \coloneq U^{\text{CNOT}}_{L-1,L} \, \cdots U^{\text{CNOT}}_{1,2} U^{\text{CNOT}}_{0,1}  = {U}_{\text{SQC}}^\dag
 \label{eq:SQC-tilde}
\end{equation}
and similarly
\begin{equation}
   \widetilde{W}_{\text{SQC}} \coloneq U^{\text{SWAP}}_{0,1} U^{\text{SWAP}}_{1,2} \, \cdots U^{\text{SWAP}}_{L-1,L}  = {W}_{\text{SQC}}^\dag
 \end{equation}
Defining superoperators analogous to Eq. (\ref{eq:definesuperoperator}), we consider the dynamics of the complementary channel involves where the $L$-qubit system is coupled to an ancilla at position $0$ in initial state $|k\rangle\langle k|$: applying the sequential circuit leaves the system in state
\begin{equation}
   \Tr_{0} \biggl[ \widetilde{\W}_{\text{SQC}} \circ \widetilde{\U}_{\text{SQC}} ( |k\rangle\langle k|  \otimes \rho )  \biggr]
\end{equation} 
analogous to Eq.~\eqref{eq:cond-state-left}.  
For later convenience, we assign the states of the ancilla with probabilities $e^{\pm s}/(e^s+e^{-s})$ and sum over these outcomes to obtain the complementary (CPTP) channel
\beq
\tM_s[\rho] \coloneq  
   \Tr_{0} \biggl[ (m_s)_0 \, \widetilde{\W}_{\text{SQC}} \circ \widetilde{\U}_{\text{SQC}} (\mathbb{1} \otimes \rho)  \biggr] \frac{1}{e^s+e^{-s}}
 \label{eq:right_channel}
\eeq
where the operator $(m_s)_0 = P_0^{(0)} e^{s} + P_0^{(1)} e^{-s}$ acts on the ancilla, the subscript indicates that this is the zeroth qubit in this case.

\subsection{Diagrammatic representation}
We introduce the standard diagrammatic representation of quantum circuits. The CNOT and SWAP gates are represented (in the unfolded representation) as
\begin{align}
    U^{\text{CNOT}} = 
    \text{
        \fineq[-0.8ex][1][1]{
            \tsfmatVthin[0][-0.5][r][1][][][blue10]
        }
    } \quad \text{and} \quad 
U^{\text{SWAP}} = 
\fineq[-0.8ex][.5][1]{
\draw[] (-0.5, 0.0)--(0.5, 2.5); 
\draw[](0.5, 0.0)--(-0.5, 2.5); 
}   
,
\end{align}
Moreover, $U^{\text{CNOT}}$ is both Hermitian and unitary so $(U^{\text{CNOT}})^2=\mathbb{1}$, that is
\begin{align}
\fineq[-0.8ex][0.8][1]{
\tsfmatVthin[0][1.25][r][1][][][blue10][topright]
\tsfmatVthin[0][0][r][1][][][blue10][topright]
\draw[ ] (-0.5,1.5) -- (-0.5,1.75);
\draw[ ] (0.5,1.5) -- (0.5,1.75);}
 =
\fineq[-0.8ex][.65][1]{
\draw[ ] (-0.45,0.15) -- (-0.45,2.35);
\draw[ ] (0.45,0.15) -- (0.45,2.35);
}.
\label{eq:CNOT-unit}
\end{align}

The operators defined by Eqs. (\ref{eq:SQC},\ref{eq:SQC-tilde}) are represented by 
\begin{align}
    U_{\text{SQC}} &= 
    \text{
        \fineq[-0.8ex][1][1]{
\tsfmatVthin[-4][3][r][1][][][blue10][topright]
\tsfmatVthin[-2][1][r][1][][][blue10][topright]
\tsfmatVthin[-1][0][r][1][][][blue10][topright]
\draw[black,dashed] (-2.75,2.75) -- (-3.25,3.25);
\draw[black, dotted] (-0.5,0.25) -- (-0.5,4.75); 
\draw[black, dotted] (-3.5,0.25) -- (-3.5,4.75); 
\draw[black, dotted] (-4.5,0.25) -- (-4.5,4.75);  
\node at (-3.5,0.15) {$i=1$};
\node at (-4.5,0.15) {$i=0$};
\node at (-0.45,0.15) {$i=L$};
}
    } \quad 
    \text{and} \quad 
    \widetilde{U}_{\text{SQC}}  &= 
    \text{
        \fineq[-0.8ex][1][1]{
\tsfmatVthin[0][1][r][1][][][blue10][topright]
\tsfmatVthin[1][2][r][1][][][blue10][topright]
\tsfmatVthin[3][4][r][1][][][blue10][topright]
\draw[black,dashed] (1.75,3.75) -- (2.25,4.25);
\draw[black, dotted] (-0.5,1.25) -- (-0.5,5.75); 
\draw[black, dotted] (0.5,1.25) -- (0.5,5.75); 
\draw[black, dotted] (3.5,1.25) -- (3.5,5.75);
\node at (0.5,1.15) {$i=1$};
\node at (3.6,1.15) {$i=L$};
\node at (-0.5,1.15) {$i=0$};
}
    }.
\end{align}

For a compact notation of quantum channels, it is convenient to work in the folded representation from quantum circuits,  see e.g.\ Ref.~\cite{bertini2025exactly}. 
To make this explicit in our notation we use a folded tensor product $\hat\otimes$: for example
\begin{align}
\U ^{\text{CNOT}} = 
 U^{\text{CNOT}} \, \hat\otimes \, U^{\text{CNOT}} =\fineq{
{\roundgatethin[0.3][1][1][topright][blue10]
\roundgatethin[0.3-.15][1 -.075][1][topright][blue10]
}
}=\fineq{\tsfmatV[0][-0.5][r][1][][][blue50]},
\end{align}
represents the unitary super-operator $\U ^{\text{CNOT}}(\cdot) =  U^{\text{CNOT}} (\cdot) U^{\text{CNOT}}$.
Then Eq.~\eqref{eq:CNOT-unit} is equivalent to
\begin{align}
 \fineq{ 
 \tsfmatV[0][-0.5][r][1][][][blue50]
\draw[fill=white,thick]  (-0.5,0) circle  (0.15);
\draw[fill=white,thick]  (0.5,0) circle  (0.15);
}  = 
\fineq{
\draw[ thick] (-0.5,0) -- (-0.5, 1);
\draw[ thick] (0.5, 0 ) -- (0.5,1);
\draw[fill=white,thick]  (-0.5,0) circle  (0.15);
\draw[fill=white,thick]  (0.5,0) circle  (0.15);
}
\quad \text{and} \quad 
 \fineq{ 
 \tsfmatV[0][-0.5][r][1][][][blue50]
\draw[fill=white,thick]  (-0.5,1) circle  (0.15);
\draw[fill=white,thick]  (0.5,1) circle  (0.15);
} = 
\fineq{
\draw[ thick] (-0.5,0) -- (-0.5, 1);
\draw[ thick] (0.5, 0 ) -- (0.5,1);
\draw[fill=white,thick]  (-0.5,1) circle  (0.15);
\draw[fill=white,thick]  (0.5,1) circle  (0.15);
},
\end{align}
where the empty circle is 
\begin{equation}
|0\rangle \,\hat\otimes\, |0\rangle + |1\rangle \,\hat\otimes\, |1\rangle
 = :  \fineq{
\draw[ thick] (0,0) -- (1, 0);
\draw[fill=white,thick]  (1,0) circle  (0.15);
}.
\end{equation}

As in the main text, we also introduce
\beq
\fineq{
\draw[ thick] (0,0) -- (1, 0);
\draw[fill=bertiniblue,draw=black, thick]  (1,0) circle  (0.15); }
\; := m_s  := \begin{pmatrix} e^{s} & 0 \\ 0 & e^{-s} \end{pmatrix} 
\label{eq:ms-sm}
\eeq
which we note may be alternatively represented [see Eq.~\eqref{eq:right_channel}] as
\beq
m_s = e^s P^{(0)} + e^{-s} P^{(1)} 
=  e^s|0\rangle \,\hat\otimes\, |0\rangle + e^{-s} |1\rangle \,\hat\otimes\, |1\rangle 
\label{eq:ms-alt}
\eeq

\section{Large Deviations and quantum Doob transform}
\label{sec:largde_deviation}
Here we review some well-established results of large deviation theory~\cite{touchette2018introduction,garrahan2018aspects,jack2020ergodicity}, and explain how to apply them to our SQC with measurement protocol.

The SQC is applied $T$ times and the measurement outcomes are recorded as $k_1,\dots,k_T$ with each $k_t$ being either $0$ or $1$.  Let $Q_0$ be the number of outcomes that are $0$; also let $Q_1$ be the number that are $1$.  Clearly $Q_0+Q_1=T$ so the full-counting statistics is fully described by the probability distribution of $\Delta Q=(Q_1-Q_0)$.  
For large times, the probability density function for $q=\Delta Q/T$ behaves as
\begin{equation}
    \text{Prob}(q) \simeq e^{-T F( q )},
\end{equation}
where $F$ is the rate function.

To make progress it is convenient to introduce a counting field $s$ such that the (unnormalised) state after a measurement outcome $k$ in Eq.~\eqref{eq:cond-state-left} is modified as
\beq
 e^{(1-2k)s}\Tr_L \biggl[  {P}^{(k)}_{L} \,   \W_{\text{SQC}} \circ \U_{\text{SQC}} (\rho \otimes \mathbb{1})  \biggr] ,
\label{eq:cond-state-left-tilt}
\eeq
This means measurements with $k=1$ are weighted by a factor $e^{-s}$ while those with $k=0$ are weighted by $e^s$.  Averaging these outcomes yields the tilted channel:
\beq
{\cal M}_s[\rho] = \sum_{k=0,1}  \Tr_L \biggl[  {P}^{(k)}_{L} e^{(1-2k)s} \, \W_{\text{SQC}} \circ \U_{\text{SQC}} (\rho \otimes \mathbb{1})  \biggr] 
\label{eq:tilted_left_channel}
\eeq

From Eq.~(\ref{eq:ms-alt}) one recognises $\sum_{k=0,1} {P}^{(k)} e^{(1-2k)s} = m_s$.  Using this fact together with the diagrammatic methods of the previous section, one sees that Eq.~\eqref{eq:tilted_left_channel} is equivalent to Eq. (\ref{eq:ms}) of the main text.  Similarly, ${\cal M}_s = ( e^s + e^{-s}) \tM_s^\dag $ which is Eq. (\ref{eq:mtilde}) of the main text (here and throughout, super-operator adjoints are defined with respect to the inner product $\langle A,B\rangle = \Tr[ A^\dag B]$ for operators $A,B$).
 
Note that the map ${\cal M}_s$ is completely positive but not trace-preserving.  Moreover, the moment generating function for $\Delta Q$ after $T$ iterations of the map ${\cal M}_s$ [starting from state $\rho_0$] is 
\beq
Z(s) = \mathbb{E}\big[e^{-s\Delta Q}\big] = \Tr\!\big( {\cal M}_s^T[\rho_0] \big)
\eeq
and one sees that the growth of $Z$ is governed by the largest eigenvalue of ${\cal M}_s$ (which is guaranteed by Perron-Frobenius theory to be real and positive): for large $T$ we have
\begin{equation}
    Z(s) \simeq e^{T \theta(s)},
\label{eq:Z-theta}
\end{equation}
with
\begin{equation}
    \theta(s) = \ln\left( \max( \text{Spec} ( \M_s ) ) \right),
\end{equation}

One also sees from Eq.~\eqref{eq:Z-theta} that $\theta$ is the scaled cumulant generating function for the large deviations of $q$, which is related to its rate function by Legendre transform, as 
\begin{equation}
    \theta(s) = -\min_{{q}}[{q}{s} + F({q})], \qquad F(q) = \max_\theta [ -qs - \theta(s) ]  \; .
\end{equation}

For ergodic channels (as considered here) the eigenmatrix of $\M_s^\dag$ associated to its dominant eigenvalue is unique and positive definite.  We denote it by $\ell_s$, that is
\beq
\M_s^\dag[\ell_s] = \ell_s e^{\theta(s)} 
\eeq
This matrix can be used for the quantum Doob transformation \cite{garrahan2010thermodynamics,cech2023thermodynamics}: one constructs a channel as
\begin{equation}
    \M_{s}^{\text{D}}[(\cdot)] = e^{-\theta(s)} \ell^{1/2}_{s} \M_{s}[\ell^{-1/2}_{s} (\cdot) \ell^{-1/2}_{s }] \ell^{1/2}_{s }
\label{eq:Doob_transformation}
\end{equation}
It may be verified that this channel is CPTP, in particular the trace-preserving property $(\M_{s}^{\text{D}})^{\dagger}[\mathbb{1}] = \mathbb{1}$ is indeed satisfied.  Moreover, rare trajectories of ${\cal M}$ become typical for $\M_{s}^{\text{D}}$, which enables efficient post-selection \cite{li2025efficient}.  It will be convenient in the following to write Eq.~\eqref{eq:Doob_transformation} as 
\beq
\M_{s}^{\text{D}} = e^{-\theta(s)} {\cal V}_s \circ \M_{s} \circ {\cal V}_s^{-1}, \quad \text{ with } \quad {\cal V}_s(\cdot) = \ell^{1/2}_{s} (\cdot)  \ell^{1/2}_{s} \; .
\label{eq:Doob_V}
\eeq

\section{Exact results for dominant eigenmatrices}
\label{sec:fixed_points}
In this section, we discuss the left and right dominant eigenmatrices of the tilted channel defined in Sec. \ref{sec:setup} and provide the derivations for the findings of the main text. 

We focus on the tilted channel Eq.~(\ref{eq:tilted_left_channel}).
Unitarity of $W_{\text{SQC}},U_{\text{SQC}}$ means that $\W_{\text{SQC}} \circ \U_{\text{SQC}}(\mathbb{1})=\mathbb{1}$ [where $\mathbb{1}$ is the identity for $(L+1)$ qubits] so the right leading dominant eigenmatrix of ${\cal M}_s$ can be straightforwardly seen to be the $L$-qubit identity:
\begin{equation}
    \M_s[ \mathbb{1}] = e^{\theta(s)} \mathbb{1},
\end{equation}
where the dominant eigenvalue is 
\begin{equation}
     e^{\theta(s)} = e^{s} + e^{-s}.
\label{eq:rate_function}
\end{equation}

The left dominant eigenmatrix is more complicated, we first define the state algebraically; then we use diagrammatic methods to show that it is an eigenstate of $\M_s^\dag$.  

We define an $L$-qubit operator
analogous to Eq.~\eqref{eq:SQC} that only acts on qubits to the right of a given location $j$:
\beq
U_{\text{SQC}}^{(j)} \coloneq U^{\text{CNOT}}_{j,j+1} \, U^{\text{CNOT}}_{j+1,j+2} \cdots U^{\text{CNOT}}_{L-2,L-1} 
\label{eq:def-USQC}
\eeq
such that $U_{\text{SQC}}^{(0)} = U_{\text{SQC}}$. The corresponding (folded) super-operator is ${\cal U}_{\text{SQC}}^{(j)} = U_{\text{SQC}}^{(j)} \,\hat\otimes\, U_{\text{SQC}}^{(j)}$.  Finally define a super-operator
\beq
{\cal U}^\triangleleft := {\cal U}_{\text{SQC}}^{(L-2)} \cdots {\cal U}_{\text{SQC}}^{(1)} \, {\cal U}_{\text{SQC}}^{(0)}
\label{eq:U-tri-sqc}
\eeq
Diagrammatically, this $L$-qubit super-operator corresponds to
\begin{align}
\label{eq:U-tri}
    \U^{\triangleleft} &= 
    \text{
        \fineq[-0.8ex][0.8][1]{
\tsfmatV[-4][3][r][1][][][blue50][topright]
\tsfmatV[-2][1][r][1][][][blue50][topright]
\tsfmatV[-1][0][r][1][][][blue50][topright]
\tsfmatV[-1][2][r][1][][][blue50][topright]
\tsfmatV[-2][5][r][1][][][blue50][topright]
\tsfmatV[-1][6][r][1][][][blue50][topright]
\tsfmatV[-1][4][r][1][][][blue50][topright]
\draw[black,dashed,thick] (-2.75,2.75) -- (-3.25,3.25);
\draw[black,dashed,thick] (-2.75,5.25) -- (-3.25,4.75);
\draw[black,dashed,thick] (-1.75,4) -- (-3.25,4);
\draw[black, dotted] (-0.5,0.25) -- (-0.5,7.75); 
\draw[black, dotted] (-3.5,0.25) -- (-3.5,7.75); 
\draw[black, dotted] (-4.5,0.25) -- (-4.5,7.75);  
\draw[ thick] (-0.5,1.5) -- (-0.5,2.5);
\draw[black, dashed,thick] (-0.5,3.65) -- (-0.5,4.35);
\draw[ thick] (-0.5,5.5) -- (-0.5,6.5);
\node at (-3.5,0.15) {$i=1$};
\node at (-4.5,0.15) {$i=0$};
\node at (-0.15,0.15) {$i=L-1$};
}
    }
\end{align}
which includes a total of $L(L-1)/2$ folded CNOT gates. 

Recalling the matrix $m_s$ defined in Eq.~\eqref{eq:ms-sm}, we show in Sec.~\ref{sec:dominant-state} below that the dominant eigenmatrix of $\M _s^{\dagger}$ is
\begin{equation}
    \rho^{\triangleleft}_s = \U^{\triangleleft} ( m_s ^{\otimes L} ),
\label{eq:triangle_state}
\end{equation}
that is,
\begin{equation}
\M _s^{\dagger}[\rho^{\triangleleft}_s] = e^{\theta(s)} \rho^{\triangleleft}_s \; .
\label{eq:left_fixed_point_eigen_equation}
\end{equation}
See Eqs.~(\ref{eq:M-fixed},\ref{eq:psi-triag}) of main text.   Physically, one sees that preparation of $\rho^{\triangleleft}_s$ requires applying a linear-depth quantum circuit on an iterated direct product of the tilting matrix $m_s$. 
Given that $m_s$ is diagonal and positive definite, one may also construct of powers of $\rho^{\triangleleft}_s$ (and its inverse), for example
\begin{equation}
\begin{aligned}
  \left(\rho^{\triangleleft}_s \right)^{1/2} &= \U^{\triangleleft}  \left( (m_s ^{1/2})^{\otimes L} \right) \\
  \left(\rho^{\triangleleft}_s \right)^{-1/2} &= \U^{\triangleleft}  \left( (m_s ^{-1/2})^{\otimes L} \right) .
\end{aligned}
\label{eq:sqrt_fixedpoint}
\end{equation}
which are useful for the Doob transformation, see below.

\subsection{{Action of ${\cal U}^{\rm CNOT}$ on projection operators}}

{The operator $U^{\rm CNOT}$ acts as a permutation in the computational basis, which is useful in the following analysis.  This fact can be exploited by noting how ${\cal U}^{\rm CNOT}$ acts on projection operators.  We first introduce the folded diagrammatic representation of projection super-operator (and operators) in the computational basis:}
\begin{equation}
\fineq{
\draw[thick] (0,0) -- (1, 0);
\draw[fill=black,thick]  (0.5,0) circle  (0.15);
\node[font=\small] at (0.5,0.4) {$n$};
}
\coloneq
\fineq{
\draw[thick] (0,0) -- (1, 0);
\draw[thick] (1.5,0) -- (2.5, 0);
\draw[fill=black,thick]  (1,0) circle  (0.15);
\draw[fill=black,thick]  (1.5,0) circle  (0.15);
\node[font=\small] at (1,0.4) {$n$};
\node[font=\small] at (1.5,0.4) {$n$};
}
= P^{(n)} \, \hat{\otimes} P^{(n)}
= |n \rangle \langle n| \hat{\otimes} |n \rangle \langle n|,
\end{equation}
for $n \in \{ 0, 1\}$. 
Note that Eq.~\eqref{eq:ms-alt} can be written as
\beq
m_s = \sum_{n\in\{0,1\}} e^{s(1-2n)} \fineq{
\draw[ thick] (0,0) -- (1, 0);
\draw[fill=black,draw=black, thick]  (1,0) circle  (0.15); \node[font=\small] at (1.4,0.1) {$n$} ; }
\label{eq:ms-n}
\eeq

From the definition of CNOT one may readily show that 
\begin{equation}
\raisebox{-3.5mm}{
\fineq{ 
\text{
\tsfmatV[0][-0.5][r][1][][][blue50]
\draw[fill=black, thick]  (-0.5,0) circle  (0.15);
\draw[fill=black, thick]  (0.5,0) circle  (0.15);
\node at (-0.5,-0.5 ) {$n_1$};
\node at (0.5, -0.5 ) {$n_2$};
}
}
}
= \quad
\fineq{
\text{
\tsfmatV[0][-0.5][r][1][][][blue50]
\draw[ thick] (-0.5, 1 ) -- (-0.75,1.25);
\draw[ thick] (0.5, 1 ) -- (0.75,1.25);
\draw[fill=black, thick]  (-0.5,0) circle  (0.15);
\draw[fill=black, thick]  (0.5,0) circle  (0.15);
\draw[fill=black, thick]  (-0.5,1) circle  (0.15);
\draw[fill=black, thick]  (0.5,1) circle  (0.15);
\node at (-0.5,-0.5 ) {$n_1$};
\node at (0.5, -0.5 ) {$n_2$};
\node at (-0.5,1.5 ) {$n_1$};
\node at (0.5, 1.5 ) {$n_1 \oplus n_2 $};
}
}
\label{eq:CNOT-prop1}
\qquad\qquad \text{{and}} \qquad\qquad
\raisebox{-2.5mm}{
 \fineq{ 
 \text{
 \tsfmatV[0][-0.5][r][1][][][blue50]
\draw[fill=black, thick]  (-0.5,0) circle  (0.15);
\draw[fill=black, thick]  (0.5,0) circle  (0.15);
\draw[fill=white, thick]  (0.5,1) circle  (0.15);
\node at (-0.5,-0.5 ) {$n_1$};
\node at (0.5, -0.5 ) {$n_2$};
} 
}
}
= \quad
\raisebox{-3mm}{
\fineq{
\text{
\draw[ thick] (-0.5,0) -- (-0.5, 1);
\draw[fill=black, thick]  (-0.5,0) circle  (0.15);
\node at (-0.5,-0.5) {$n_1$};
}
}
},
\end{equation}
Specifically, the left output of CNOT is unaffected by its right input. Combining these two relations also implies
\begin{align}
\begin{split}
\fineq{
\tsfmatV[1][0.5][r][1][][][blue50]
\tsfmatV[1][-1.5][r][1][][][blue50]
\draw[thick] (1.5,0.) -- (1.5, 1);
\draw[thick] (0.5,-1) -- (0.25 , -1.25);
\draw[thick] (1.5,-1.) -- (1.75, -1.25);
\draw[thick] (0.5,1.) -- (0.25, 0.75);
\draw[fill=black, thick]  (0.5,1) circle  (0.15);
\draw[fill=black, thick]  (0.5,-1) circle  (0.15);
\draw[fill=black, thick]  (1.5,-1) circle  (0.15);
\draw[fill=white, thick]  (1.5,2) circle  (0.15);
\node at (0,1) {$n_3$};
\node at (0,-1) {$n_1$};
\node at (2,-1) {$n_2$};
} 
\quad = \quad
\fineq{
\tsfmatV[1][-1.5][r][1][][][blue50]
\draw[thick] (0.5,-1) -- (0.25 , -1.25);
\draw[thick] (1.5,-1.) -- (1.75, -1.25);
\draw[thick] (0.5,1.5) -- (0.5, 0.5);
\draw[fill=black, thick]  (0.5,1) circle  (0.15);
\draw[fill=black, thick]  (0.5,-1) circle  (0.15);
\draw[fill=black, thick]  (1.5,-1) circle  (0.15);
\draw[fill=white, thick]  (1.5,0) circle  (0.15);
\node at (0,1) {$n_3$};
\node at (0,-1) {$n_1$};
\node at (2,-1) {$n_2$};
} .
\label{eq:CNOT-prop3}
\end{split}
\end{align}
which will be used extensively in the following.

\subsection{Dominant eigenmatrix of $\M^\dag_s$}
\label{sec:dominant-state}

To derive Eq.~\eqref{eq:left_fixed_point_eigen_equation} we note [by Eqs.~(\ref{eq:tilted_left_channel},\ref{eq:U-tri},\ref{eq:triangle_state},\ref{eq:ms-n})] that 
\beq
\M _s^{\dagger}[\rho^{\triangleleft}_s] = \sum_{ k, n_0, \dots, n_{L-1} } {\rm e}^{\sum_i (1- 2 n_i)s}  f^\triangleleft(k,n_0,\dots,n_{L-1})
\label{eq:M-f}
\eeq
where all sums run over the set $\{0,1\}$ and
\beq
f^\triangleleft(k,n_0,\dots,n_{L-1}) = \text{
\fineq[-0.8ex][0.8][1]{ 
\tsfmatV[-4][3][r][1][][][blue50][topright]
\tsfmatV[-2][1][r][1][][][blue50][topright]
\tsfmatV[-1][0][r][1][][][blue50][topright]
\tsfmatV[-1][2][r][1][][][blue50][topright]
\tsfmatV[-2][5][r][1][][][blue50][topright]
\tsfmatV[-1][6][r][1][][][blue50][topright]
\tsfmatV[-1][4][r][1][][][blue50][topright]
\draw[black,dashed,thick] (-2.75,2.75) -- (-3.25,3.25);
\draw[black,dashed,thick] (-2.75,5.25) -- (-3.25,4.75);
\draw[black,dashed,thick] (-1.75,4) -- (-3.25,4);
\draw[ thick] (-0.5,1.5) -- (-0.5,2.5);
\draw[black, dashed,thick] (-0.5,3.65) -- (-0.5,4.35);
\draw[ thick] (-0.5,5.5) -- (-0.5,6.5);
\tsfmatV[-5][4][r][1][][][blue50][topright]
\tsfmatV[-3][6][r][1][][][blue50][topright]
\tsfmatV[-2][7][r][1][][][blue50][topright]
\tsfmatV[-1][8][r][1][][][blue50][topright]
\draw[ thick] (-0.5,7.5) -- (-0.5,8.5);
\draw[black,dashed,thick] (-3.75, 6.25) -- (-4.25, 5.75);
\draw[fill=white, thick]  (-0.5,9.5) circle  (0.15);
\draw[fill=black, thick]  (-0.5,0.5) circle  (0.15);
\draw[fill=black, thick]  (-1.5,0.5) circle  (0.15);
\draw[fill=black, thick]  (-2.5,1.5) circle  (0.15);
\draw[fill=black, thick]  (-4.5,3.5) circle  (0.15);
\draw[fill=black, thick]  (-5.5,4.5) circle  (0.15);
\node at (-0.5,0 ) {$n_{L-1}$};
\node at (-1.5,0) {$n_{L-2}$};
\node at (-2.5,1) {$n_{L-3}$};
\node at (-4.5,3) {$n_0$};
\node at (-5.5,4) {$k$};
}
}
\eeq
Hence we have
\begin{align}
f^\triangleleft(k,n_0,\dots,n_{L-1})  = 
\text{
\fineq[-0.8ex][0.8][1]{  
\tsfmatV[-4][3][r][1][][][blue50][topright]
\tsfmatV[-2][1][r][1][][][blue50][topright]
\tsfmatV[-1][0][r][1][][][blue50][topright]
\tsfmatV[-1][2][r][1][][][blue50][topright]
\tsfmatV[-2][5][r][1][][][blue50][topright]
\tsfmatV[-1][6][r][1][][][blue50][topright]
\tsfmatV[-1][4][r][1][][][blue50][topright]
\draw[black,dashed,thick] (-2.75,2.75) -- (-3.25,3.25);
\draw[black,dashed,thick] (-2.75,5.25) -- (-3.25,4.75);
\draw[black,dashed,thick] (-1.75,4) -- (-3.25,4);
\draw[ thick] (-0.5,1.5) -- (-0.5,2.5);
\draw[black, dashed,thick] (-0.5,3.65) -- (-0.5,4.35);
\draw[ thick] (-0.5,5.5) -- (-0.5,6.5);
\tsfmatV[-5][4][r][1][][][blue50][topright]
\tsfmatV[-3][6][r][1][][][blue50][topright]
\tsfmatV[-2][7][r][1][][][blue50][topright]
\tsfmatV[-1][8][r][1][][][blue50][topright]
\draw[ thick] (-0.5,7.5) -- (-0.5,8.5);
\draw[black,dashed,thick] (-3.75, 6.25) -- (-4.25, 5.75);
\draw[fill=white, thick]  (-0.5,9.5) circle  (0.15);
\draw[fill=black, thick]  (-0.5,0.5) circle  (0.15);
\draw[fill=black, thick]  (-1.5,0.5) circle  (0.15);
\draw[fill=black, thick]  (-2.5,1.5) circle  (0.15);
\draw[fill=black, thick]  (-4.5,3.5) circle  (0.15);
\draw[fill=black, thick]  (-5.5,4.5) circle  (0.15);
\draw[fill=black, thick]  (-0.5,2) circle  (0.15);
\draw[fill=black, thick]  (-0.5,6) circle  (0.15);
\draw[fill=black, thick]  (-0.5,8) circle  (0.15);
\draw[fill=black, thick]  (-1.5,1.5) circle  (0.15);
\draw[fill=black, thick]  (-1.5,2.5) circle  (0.15);
\draw[fill=black, thick]  (-1.5,5.5) circle  (0.15);
\draw[fill=black, thick]  (-1.5,6.5) circle  (0.15);
\draw[fill=black, thick]  (-1.5,7.5) circle  (0.15);
\draw[fill=black, thick]  (-1.5,8.5) circle  (0.15);
\node at (-0.5,0 ) {$n_{L-1}$};
\node at (-1.5,0) {$n_{L-2}$};
\node at (-2.5,1) {$n_{L-3}$};
\node at (-4.5,3) {$n_0$};
\node at (-5.5,4) {$k$};
}
} 
=  
\text{
\fineq[-0.8ex][0.8][1]{  
\tsfmatV[-4][3][r][1][][][blue50][topright]
\tsfmatV[-2][1][r][1][][][blue50][topright]
\tsfmatV[-2][5][r][1][][][blue50][topright]
\draw[black,dashed,thick] (-2.75,2.75) -- (-3.25,3.25);
\draw[black,dashed,thick] (-2.75,5.25) -- (-3.25,4.75);
\draw[black,dashed,thick] (-2,4) -- (-3,4);
\draw[ thick] (-1.5,2.5) -- (-1.5,3.5);
\draw[ thick] (-1.5,4.5) -- (-1.5,5.5);
\draw[black, dashed,thick] (-1.5,3.65) -- (-1.5,4.35);
\tsfmatV[-5][4][r][1][][][blue50][topright]
\tsfmatV[-3][6][r][1][][][blue50][topright]
\tsfmatV[-2][7][r][1][][][blue50][topright]
\draw[ thick] (-1.5,6.5) -- (-1.5,7.5);
\draw[black,dashed,thick] (-3.75, 6.25) -- (-4.25, 5.75);
\draw[fill=black, thick]  (-1.5,1.5) circle  (0.15);
\draw[fill=black, thick]  (-2.5,1.5) circle  (0.15);
\draw[fill=black, thick]  (-4.5,3.5) circle  (0.15);
\draw[fill=black, thick]  (-5.5,4.5) circle  (0.15);
\node at (-1.5,1) {$n_{L-2}$};
\node at (-2.5,1) {$n_{L-3}$};
\node at (-4.5,3) {$n_0$};
\node at (-5.5,4) {$k$};
}
} 
\end{align}
where the first equality uses repeatedly Eq.~\eqref{eq:CNOT-prop1} and the second uses Eq.~\eqref{eq:CNOT-prop3} repeatedly to remove $L$ gates and then uses again Eq.~\eqref{eq:CNOT-prop1}; black dots without labels refer to projection operators on states whose specific value is not needed for the derivation.
Plugging this last result back into Eq.~\eqref{eq:M-f}, the sums can be performed and the result is exactly Eq.~\eqref{eq:left_fixed_point_eigen_equation}.

This diagrammatic computation somewhat resembles those of 
Refs. \cite{gopalakrishnan2019unitary,giudice2022temporal} but the results here are different and rely on specific properties of CNOT gates.

\section{quantum Doob transformation}
\label{sec:Doob}

Using  Eq.~\eqref{eq:Doob_V} for the Doob transform of $\tilde \M_s$ together with Eq.~\eqref{eq:left_fixed_point_eigen_equation}, we identify 
$  \V_s(\cdot) 
    = 
    (\rho^{\triangleleft}_s)^{1/2} (\cdot) (\rho^{\triangleleft}_s)^{1/2} 
    $
    and so
    \beq
      \V_s = e^{-L\theta(s)} \U^\triangleleft \circ (m_s^{1/2} {\hat \otimes} \, m_s^{1/2} )^{\otimes L} \circ \U^\triangleleft
    \,.
\eeq
Similarly $\V_s^{-1} = e^{L\theta(s)} \U^\triangleleft \circ (m_s^{-1/2} {\hat \otimes} \, m_s^{-1/2} )^{\otimes L} \circ \U^\triangleleft$. This allows $\M^{\text{D}}_{{s}}$ to be constructed as in Fig.~\ref{fig:fig2} of the main text, leading to a linear-depth quantum circuit. Explicitly writing
\beq
\M^{\text{D}}_{{s}} = e^{-\theta(s)}
\U^\triangleleft \circ (m_s^{1/2} {\hat \otimes} \, m_s^{1/2} )^{\otimes L} \circ \U^\triangleleft
\circ \M_s \circ 
\U^\triangleleft \circ (m_s^{-1/2} {\hat \otimes} \, m_s^{-1/2} )^{\otimes L} \circ \U^\triangleleft
\label{eq:doob-explicit}
\eeq
shows the presence of four $\U^\triangleleft$'s (triangles) as well as $L$ instances of $(m_s^{1/2} {\hat \otimes} \, m_s^{1/2} )$ and $L$ of $(m_s^{-1/2} {\hat \otimes} \, m_s^{-1/2} )$.

Since $\U^\triangleleft$ is a linear-depth many-body circuit, this Doob transformation is not locality preserving, i.e., it can map local operators to non-local ones.  This property arses from the long-range correlations in $\rho^\triangleleft$.
In the past, explicit forms of the (non-trivial) Doob transformation were only explored in limited systems \cite{popkov2010asep} or systems in the hydrodynamic limit \cite{jack2015hyperuniformity}. This is because performing the Doob transformation involves a number of challenging steps, including the explicit calculations of the rate function and the left dominant eigenmatrix of a many-body system, as well as taking the matrix square root and inverse of the left dominant eigenmatrix \cite{PhysRevE.100.020103}. Thanks to the linear-depth circuit structure of the left dominant eigenmatrix of $\M_s$ in Eq. (\ref{eq:triangle_state}), Eq.~\eqref{eq:doob-explicit} provides the first example of a non-trivial many-body Doob channel for arbitrary system sizes, in explicit form.

\section{Inhomogeneous tilting }
\label{sec:Inhomogeneous}

As discussed in the main text, the spectrum of the tilted channel ${\cal M}_s$ features a Jordan block structure so that the system reaches the steady state exactly in $L$ time steps.  This structure also allows analysis of a channel where the value of the tilting field $s$ is different on each time step (inhomogeneous tilt).

The inhomogeneous tilted channel for $t$ time steps is
\beq
\M_{\{s_t\}} = \M_{s_{t}} \circ \dots \circ \M_{s_2} \circ  \M_{s_1}
\eeq
It is convenient to work with the adjoint of this superoperator (note the time-ordering): $\M_{\{s_t\}}^\dag = \M_{s_{1}}^\dag \circ \dots \circ \M_{s_{t-1}}^\dag \circ  \M_{s_t}^\dag$.  The corresponding complementary channel is 
\beq
\tM_{\{s_t\}} = \tM_{s_{t}} \circ \dots \circ \tM_{s_2} \circ  \tM_{s_1}
\label{eq:tM-compo}
\eeq
One may verify that this is related to $\M_{\{s_t\}}^\dag$ by a multiplicative factor and a re-ordering of the time indices.

\subsection{Diagonal initial states}

Now use  Eq.~\eqref{eq:ms-n} to obtain the action of this $t$-fold channel on an arbitrary diagonal basis state of the computational basis:
\begin{equation}
    \M_{\{ s_{t} \}}^\dag\!\left(P^{(n_0)} \otimes \dots \otimes P^{(n_{L-1})}\right) = \sum_{ w_1,\dots,w_t }  e^{\sum_{i} (1-2w_{i})s_i} g^\square(w_1,\dots,w_t;n_0,\dots,n_{L-1})
    \label{eq:W-g-square}
\end{equation}
where each variable $w_i$ is summed over $\{0,1\}$ and similarly every $n_i\in\{0,1\}$; also 
\begin{align}
 g^\square(w_1,\dots,w_t;n_0,\dots,n_{L-1})
 & =
\text{
\fineq[-0.8ex][0.8][1]{  
\tsfmatV[-4][3][r][1][][][blue50][topright]
\tsfmatV[-2][1][r][1][][][blue50][topright]
\tsfmatV[-1][0][r][1][][][blue50][topright]
\tsfmatV[-1-1][6-1][r][2][][][blue50][topright]
\tsfmatV[+1-1][4-1][r][2][][][blue50][topright]
\tsfmatV[+2-1][3-1][r][2][][][blue50][topright]
\tsfmatV[0][7][r][1][][][blue50][topright]
\tsfmatV[2][5][r][1][][][blue50][topright]
\tsfmatV[3][4][r][1][][][blue50][topright]
\tsfmatV[-5][4][r][1][][][blue50][topright]
\tsfmatV[-3][6][r][1][][][blue50][topright]
\tsfmatV[-2][7][r][1][][][blue50][topright]
\tsfmatV[-1][8][r][1][][][blue50][topright]
\draw[black,dashed,thick] (-2.75 ,2.75) -- (-3.25,3.25);
\draw[black,dashed,thick] (-1.75 +1 ,3.75 +1) -- (-2.25 + 1,4.25 +1);
\draw[black,dashed,thick] (2+-1.75 ,2+3.75) -- (2+-2.25,2+4.25);
\draw[black,dashed,thick] (2+-1.75+1 ,2+3.75+1) -- (2+-2.25+1,2+4.25+1);
\draw[black,dashed,thick] (0.5-1,2.5-1) -- (1.5-1,3.5-1);
\draw[black,dashed,thick] (0.5- 1 -1 ,2.5 + 1 -1) -- (1.5 - 1 -1,3.5 + 1-1);
\draw[black,dashed,thick] (0.5-3 -1,2.5 + 3-1) -- (1.5 - 3 -1 ,3.5 + 3 -1);
\draw[black,dashed,thick] (0.5-3 -1-1,2.5 + 3-1+1) -- (1.5 - 3 -1 -1 ,3.5 + 3 -1 +1);
\draw[fill=black, thick]  (-0.5,0.5) circle  (0.15);
\draw[fill=black, thick]  (1.5,2.5) circle  (0.15);
\draw[fill=black, thick]  (2.5,3.5) circle  (0.15);
\draw[fill=black, thick]  (3.5,4.5) circle  (0.15);
\draw[fill=black, thick]  (-1.5,0.5) circle  (0.15);
\draw[fill=black, thick]  (-2.5,1.5) circle  (0.15);
\draw[fill=black, thick]  (-4.5,3.5) circle  (0.15);
\draw[fill=black, thick]  (-5.5,4.5) circle  (0.15);
\draw[fill=white, thick]  (3.5,5.5) circle  (0.15);
\draw[fill=white, thick]  (2.5,6.5) circle  (0.15);
\draw[fill=white, thick]  (0.5,8.5) circle  (0.15);
\draw[fill=white, thick]  (-0.5,9.5) circle  (0.15);
\node at (-0.5,0.0) {$n_0$};
\node at (1.5,2) {$n_{L-3}$};
\node at (2.5,3) {$n_{L-2}$};
\node at (3.5,4) {$n_{L-1}$};
\node at (-1.5,0.0) {$w_{t}$};
\node at (-2.5,1) {$w_{t-1}$};
\node at (-4.5,3) {$w_{2}$};
\node at (-5.5,4) {$w_1$};
}
}
\end{align}

For $t\geq L$ one uses Eqs. (\ref{eq:CNOT-prop1}) and (\ref{eq:CNOT-prop3}) to reduce this to
\begin{align}
 g^\square(w_1,\dots,w_t;n_0,\dots,n_{L-1})
 =
\text{
\fineq[-0.8ex][0.8][1]{
\tsfmatV[-4][3][r][1][][][blue50][topright]
\tsfmatV[-2][1][r][1][][][blue50][topright]
\tsfmatV[-2][5][r][1][][][blue50][topright]
\draw[black,dashed,thick] (-2.75,2.75) -- (-3.25,3.25);
\draw[black,dashed,thick] (-2.75,5.25) -- (-3.25,4.75);
\draw[black,dashed,thick] (-2,4) -- (-3,4);
\draw[ thick] (-1.5,2.5) -- (-1.5,3.5);
\draw[ thick] (-1.5,4.5) -- (-1.5,5.5);
\draw[black, dashed,thick] (-1.5,3.65) -- (-1.5,4.35);
\tsfmatV[-5][4][r][1][][][blue50][topright]
\tsfmatV[-3][6][r][1][][][blue50][topright]
\tsfmatV[-2][7][r][1][][][blue50][topright]
\draw[ thick] (-1.5,6.5) -- (-1.5,7.5);
\draw[black,dashed,thick] (-3.75, 6.25) -- (-4.25, 5.75);
\draw[fill=black, thick]  (-1.5,1.5) circle  (0.15);
\draw[fill=black, thick]  (-2.5,1.5) circle  (0.15);
\draw[fill=black, thick]  (-4.5,3.5) circle  (0.15);
\draw[fill=black, thick]  (-5.5,4.5) circle  (0.15);
\node at (-1.2,1) {$w_{L}$};
\node at (-2.8,1) {$w_{L-1}$};
\node at (-4.5,3) {$w_{2}$};
\node at (-5.5,4) {$w_1$};
}
} ,
\label{eq:g-becomes-triag}
\end{align}
which we notice is fully determined by $w_{1},\dots,w_L$; it is independent of the initial state $(n_0,\dots,n_{L-1})$ and of $(w_{L+1},\dots,w_{t})$.  Noting the similarity with Eq.~\eqref{eq:U-tri} and using again Eq.~\eqref{eq:ms-n}, one has from Eq.~\eqref{eq:W-g-square} that 
\beq
\M_{\{ s_{t} \}}^\dag\!\left(P^{(n_0)} \otimes \dots \otimes P^{(n_{L-1})}\right) = {\cal U}^\triangleleft\!\left( m_1\otimes \cdots \otimes m_{L} \right) \prod_{t'=1}^L e^{\theta(s_{t'})}
\label{eq:Mst-diag}
\eeq
Note that this depends on $s_{1},\dots,s_L$ via the matrices $m_{1},\dots,m_L$ but is independent of $n_0,\dots,n_{L-1}$ (as noted above).

\subsection{Off-diagonal initial states}

To characterise in more detail the properties of $\M_{\{s_t\}}^\dag$ we note that Eq.~\eqref{eq:Mst-diag} gives its action on diagonal basis states; we now consider its action on off-diagonal basis states. It is useful to define a folded unitary circuit, acting on a system of $t + L$ qubits, 
\begin{align}
\label{eq:U-sq}
\U^{\square} = U^{\square} \hat{\otimes} U^{\square} \coloneq 
\text{
\fineq[-0.8ex][0.8][1]{
\tsfmatV[-4][3][r][1][][][blue50][topright]
\tsfmatV[-2][1][r][1][][][blue50][topright]
\tsfmatV[-1][0][r][1][][][blue50][topright]
\tsfmatV[-1-1][6-1][r][2][][][blue50][topright]
\tsfmatV[+1-1][4-1][r][2][][][blue50][topright]
\tsfmatV[+2-1][3-1][r][2][][][blue50][topright]
\tsfmatV[0][7][r][1][][][blue50][topright]
\tsfmatV[2][5][r][1][][][blue50][topright]
\tsfmatV[3][4][r][1][][][blue50][topright]
\tsfmatV[-5][4][r][1][][][blue50][topright]
\tsfmatV[-3][6][r][1][][][blue50][topright]
\tsfmatV[-2][7][r][1][][][blue50][topright]
\tsfmatV[-1][8][r][1][][][blue50][topright]
\draw[black,dashed,thick] (-2.75 ,2.75) -- (-3.25,3.25);
\draw[black,dashed,thick] (-1.75 +1 ,3.75 +1) -- (-2.25 + 1,4.25 +1);
\draw[black,dashed,thick] (2+-1.75 ,2+3.75) -- (2+-2.25,2+4.25);
\draw[black,dashed,thick] (2+-1.75+1 ,2+3.75+1) -- (2+-2.25+1,2+4.25+1);
\draw[black,dashed,thick] (0.5-1,2.5-1) -- (1.5-1,3.5-1);
\draw[black,dashed,thick] (0.5- 1 -1 ,2.5 + 1 -1) -- (1.5 - 1 -1,3.5 + 1-1);
\draw[black,dashed,thick] (0.5-3 -1,2.5 + 3-1) -- (1.5 - 3 -1 ,3.5 + 3 -1);
\draw[black,dashed,thick] (0.5-3 -1-1,2.5 + 3-1+1) -- (1.5 - 3 -1 -1 ,3.5 + 3 -1 +1);
\draw[<->, thick] (0,0) -- (4,4);
\draw[<->, thick] (-2,0) -- (-6,4);
\node at (2+0.5, 1.5) {$L$};
\node at (-4-0.5, 1.5) {$t$};
}
}.
\end{align}
where $U^\square$ is the corresponding unfolded circuit.  
The CNOT gate acts as a permutation on computational basis states and this property is inherited by $U^\square$.  This means that
\begin{equation}
    U^{\square} \big( | w_{t},\dots w_{1} \rangle \otimes |n_{0} , \dots n_{L-1} \rangle \big) =  | q_{0},\dots q_{L-1} \rangle \otimes |b_{1} , \dots b_{t} \rangle 
    \label{eq:U-wnqb}
\end{equation}
where $q_0,\dots,q_{L-1},b_1,\dots,b_t$ are binary-valued functions of the binary variables $w_1,\dots,w_t,n_0,\dots,n_{L-1}$; note the circuit is rectangular in general so the partitioning of the variables into groups is different for input and output. Applying $\M_{\{ s_{t} \}} ^{\dagger}$ to an arbitrary off-diagonal basis state, one obtains
\begin{align}
\M_{\{ s_{t} \}} ^{\dagger} \left( |n\rangle \langle n' | \right)  &=  \sum_{w_1,\dots,w_t} e^{\sum (1-2 w_i) s_i}
\text{
\fineq[-0.8ex][0.8][1]{
\tsfmatV[-4][3][r][1][][][blue50][topright]
\tsfmatV[-2][1][r][1][][][blue50][topright]
\tsfmatV[-1][0][r][1][][][blue50][topright]
\tsfmatV[-1-1][6-1][r][2][][][blue50][topright]
\tsfmatV[+1-1][4-1][r][2][][][blue50][topright]
\tsfmatV[+2-1][3-1][r][2][][][blue50][topright]
\tsfmatV[0][7][r][1][][][blue50][topright]
\tsfmatV[2][5][r][1][][][blue50][topright]
\tsfmatV[3][4][r][1][][][blue50][topright]
\tsfmatV[-5][4][r][1][][][blue50][topright]
\tsfmatV[-3][6][r][1][][][blue50][topright]
\tsfmatV[-2][7][r][1][][][blue50][topright]
\tsfmatV[-1][8][r][1][][][blue50][topright]
\draw[black,dashed,thick] (-2.75 ,2.75) -- (-3.25,3.25);
\draw[black,dashed,thick] (-1.75 +1 ,3.75 +1) -- (-2.25 + 1,4.25 +1);
\draw[black,dashed,thick] (2+-1.75 ,2+3.75) -- (2+-2.25,2+4.25);
\draw[black,dashed,thick] (2+-1.75+1 ,2+3.75+1) -- (2+-2.25+1,2+4.25+1);
\draw[black,dashed,thick] (0.5-1,2.5-1) -- (1.5-1,3.5-1);
\draw[black,dashed,thick] (0.5- 1 -1 ,2.5 + 1 -1) -- (1.5 - 1 -1,3.5 + 1-1);
\draw[black,dashed,thick] (0.5-3 -1,2.5 + 3-1) -- (1.5 - 3 -1 ,3.5 + 3 -1);
\draw[black,dashed,thick] (0.5-3 -1-1,2.5 + 3-1+1) -- (1.5 - 3 -1 -1 ,3.5 + 3 -1 +1);
\draw[fill=black, thick]  (-1.5,0.5) circle  (0.15);
\draw[fill=black, thick]  (-2.5,1.5) circle  (0.15);
\draw[fill=black, thick]  (-4.5,3.5) circle  (0.15);
\draw[fill=black, thick]  (-5.5,4.5) circle  (0.15);
\draw[fill=white, thick]  (3.5,5.5) circle  (0.15);
\draw[fill=white, thick]  (2.5,6.5) circle  (0.15);
\draw[fill=white, thick]  (0.5,8.5) circle  (0.15);
\draw[fill=white, thick]  (-0.5,9.5) circle  (0.15);
\node at (-0.5 + 0.1,0.2) {$|n_0\rangle\langle n'_{0}|$};
\node at (1.5 + 0.1,2.2) {$|n_{L-3}\rangle\langle n'_{L-3}|$};
\node at (2.5 + 0.1,3.2) {$|n_{L-2}\rangle\langle n'_{L-2}|$};
\node at (3.5 + 0.1,4.2) {$|n_{L-1}\rangle\langle n'_{L-1}|$};
\node at (-1.5,0.0) {$w_{1}$};
\node at (-2.5,1) {$w_{2}$};
\node at (-4.5,3) {$w_{t-1}$};
\node at (-5.5,4) {$w_t$};
}
} 
\label{eq:M-sq}
\end{align}
Recalling \eqref{eq:U-sq} and writing Eq.~\eqref{eq:U-wnqb} as $U^{\square} \big( | w \rangle \otimes |n \rangle \big) =  | q(w,n) \rangle \otimes |b(w,n) \rangle $,
we find
\beq
\M_{\{ s_{t} \}} ^{\dagger} \left( |n\rangle \langle n' | \right)  =   \sum_{w_1,\dots,w_t} e^{\sum_i (1-2 w_i) s_i} | q(w,n) \rangle \langle q(w,n') | \, \langle  b(w,n) |  b(w,n') \rangle
\eeq

We now show that this vanishes for $t\geq L$ and $n\neq n'$ which means that $\M_{\{ s_{t} \}} ^{\dagger} ( \rho )$ is diagonal for $t \geq L$, as asserted above.  Using Eq.~(\ref{eq:CNOT-prop1}) with Eqs.~(\ref{eq:U-sq},\ref{eq:U-wnqb}) one sees that if $t \geq L$, then $q(w,n)$ is fully determined by $w_1,\dots,w_t$; in particular it is independent of $n$.  Since $U^\square$ is one-to-one (in fact, unitary), this means that $b(w,n')\neq b(w,n)$ unless $n=n'$.  Hence for $t \geq L$ 
\beq
\M_{\{ s_{t} \}} ^{\dagger} \left( |n\rangle \langle n' | \right)  =   \sum_{w_1,\dots,w_t} e^{\sum_i (1-2 w_i) s_i} | q(w,n) \rangle \langle q(w,n) | \delta_{n,n'}
\label{eq:Mst-delta}
\eeq
which vanishes for off-diagonal initial states, as required.  [For $t<L$ and $n\neq n'$ it can be that $b(w,n)= b(w,n')$ because $q(w,n),q(w,n')$ depend on $n,n'$ and may therefore be different.]
Eq.~\eqref{eq:Mst-delta} is consistent with Eq.~\eqref{eq:Mst-diag} on taking $n=n'$; it additionally shows that coherences between computational basis states are destroyed in $L$ steps.

\subsection{Spectrum of $\M_s$ (Jordan blocks)}
\label{sec:jordan-m}

The results above also establish the Jordan block structure of $\M_s$.  Specifically Eqs.~\eqref{eq:Mst-diag} and~\eqref{eq:Mst-delta} show that for $t\geq L$ then $\M_{\{s_t\}}^\dag$ has a single positive eigenvalue and $2^{2L}-1$ zeros.  For the special case where $s_1=s_2=\dots=s_t=s$, this means that $\M_s^\dag$ has a unique non-zero eigenvalue of $e^{\theta(s)}$, but it may generally contain Jordan blocks with sizes less than or equal to $L$.  An argument similar to that of Eq.~\eqref{eq:g-becomes-triag} shows that for $t<L$ the state $ g^\square(w_1,\dots,w_t;n_0,\dots,n_{L-1})$ does depend on $n_0$; it follows that the largest Jordan block of $\M_s^\dag$ has size exactly $L$.  The spectrum of $\M_s$ (and the sizes of its Jordan blocks) are the same as those of $\M_s^\dag$.

\subsection{Complementary channel $\tM_s$}

Starting from Eq.~\eqref{eq:tM-compo} one also sees [analogous to Eq.~\eqref{eq:W-g-square}] that 
\beq
    \tM_{\{ s_{t} \}} \!\left(P^{(n_0)} \otimes \dots \otimes P^{(n_{L-1})}\right) = \sum_{ w_1,\dots,w_t }  e^{\sum_{i} [(1-2w_{i})s_i - \theta(s_i)]} g^\square(w_t,\dots,w_1;n_0,\dots,n_{L-1})
    \label{eq:tM-g}
\eeq
Using this formula and repeating the argument leading to Eq.~\eqref{eq:Mst-diag} yields Eq.~\eqref{eq:rtL} of the main text.
Similarly, a result analogous to Eq. \eqref{eq:Mst-delta} shows that $\tM_{\{s_t\}}$ destroys all coherences in $L$ steps.  Combined with the argument of Sec.~\ref{sec:jordan-m}, this is sufficient for the result given in the main text: that the single-step channel $\tM_s$ has a single non-zero eigenvalue (which is unity); all other eigenvalues are zero but there are also Jordan blocks with maximal size $L$.

\section{Observables}
\label{sec:Observables}

In this section, we provide the full details of the two-point correlator calculations discussed in the main text. In Sec. \ref{sec:fixed_points}, we have shown that any non-trivial tilting leads to a highly long-range correlated density matrix Eq. (\ref{eq:triangle_state}). Here, we explore how long-range correlations of the dominant eigenmatrices are reflected at the level of physical observables. The underlying (doubled) Hilbert space is spanned by all possible Pauli-strings, which provides a formal way of expressing any operators in terms of a complete basis, e.g., the density matrices $\rho \in \text{End}(\mathcal{H}^{L} \otimes \mathcal{H}^{L} )$. Any traceless observables $O \in \text{End}(\mathcal{H}^{L} \otimes \mathcal{H}^{L} )$ can be expanded in the Pauli-strings basis. We would like to compute the expectation values of observables for the tilted protocol.

For a general tilted CP map $\M_{s}$, we define the tilted observables as following 
\begin{equation}
    \langle O(t) \rangle_{\rm s} \coloneq \frac{ \Tr\left( \M_{ s}^{T-t}\! \left[ O \M_{s}^{t}[\rho_0]\right] \right) }{ \Tr( \M_{ s}^{T}[\rho_0]) } = \frac{\Tr[ l_{\bold{s}} O r_{\bold{s}} ]}{\Tr[ l_{\bold{s}} r_{\bold{s}} ]},
\label{eq:observable}
\end{equation}
to avoid transits at the short and late times. {As shown in the previous section, the Jordan block structure of the tilted channel guarantees any initial density matrix is relaxed to the dominant eigenmatrix for any $t \ge L$. Hence, the second equality holds for $L \le t \le T-L$. }

To be concrete with the focus of this work, in the case of $\M_{s}$, and right dominant eigenmatrix is trivial $r_{s}  = \mathbb{1}^{\otimes L}$. Thus, the evaluation of $\langle O(L \le t \le T-L) \rangle_{\rm s}$ in Eq. (\ref{eq:observable}) becomes 
\begin{equation}
   \langle O \rangle_s =  \frac{\Tr[ O \rho^{\triangleleft}_s ]}{\Tr[\rho^{\triangleleft}_s ]} = \frac{\Tr[ O \rho^{\triangleleft}_s ]}{\Tr[ m_{s}^{\otimes L} ]}  \coloneq \frac{\Tr[ \tilde{O} m_{s} ^{\otimes L} ]}{\Tr[ m_{s}^{\otimes L} ]} ,
\label{eq:observable_CNOT}
\end{equation}
where we have introduced 
\begin{equation}
    \tilde{O} \coloneq \U^{\triangleleft} (O ),
\label{eq:tilde_U}
\end{equation}
and the normalization of $\rho^{\triangleleft}_s$ is given by $\Tr( \rho^{\triangleleft}_s) = \Tr[ m_{s}^{\otimes L} ] = (e^{s} + e^{-s})^{L}$.  
Note that if $\tilde O$ is a Pauli string then these observables can be straightforwardly evaluated, in particular $m_s$ is diagonal so $\operatorname{Tr}[ \tilde{O} m_{s} ^{\otimes L} ]=0$ if $\tilde{O}$ contains any $X$s or $Y$s; for Z strings we have
\begin{equation}
    \frac{\Tr[ \left( \bigotimes_{i = 0}^{L-1} Z^{z_{i}} \right) m_{s}^{\otimes L} ]}{\Tr[m_{s}^{\otimes L}]} 
    = \left( \frac{ e^{s} - e^{-s} }{ e^{s}  + e^{-s}}\right)^{\sum_i z_i}
    \; .
\label{eq:Obs_Z_string}
\end{equation}

In the following, we characterise $\tilde{O}$ for all possible Pauli-strings $O$. 
Since CNOT belongs to the family of \textit{Clifford} gates and is unitary, $\tilde{O}$ must also be a Pauli string. Knowing $\tilde{O}$ for strings formed of $Z$s and $X$s is enough to construct all the possible Pauli-strings as $i Y = Z X $. Let us first consider the case of $Z$ observables, and the most general $Z$ Pauli-string is given by $\bigotimes_{i = 0}^{L-1} Z_{i}^{z_{i}} $, where $z_{i} \in \{ 0 , 1\}$, and $Z_i$ acts as Pauli-$Z$ on the $i$th qubit and as identity elsewhere. The string $(z_0,z_1,\dots,z_{L-1})$ specifies the locations of $Z$ operators. We will show that
\begin{equation}
    \U^{(0)}_{\text{SQC}}
       \left( \bigotimes_{i = 0}^{L-1} Z^{z_{i}} \right) 
     = \bigotimes_{i = 0}^{L-1} Z^{\oplus_{j = i }^{L-1}z_{j}},
\label{eq:Z_paulistring}
\end{equation}
where $\oplus$ indicates the XOR operation (that is, addition modulo 2, and not the direct sum).

Using the diagrammatic representation introduced in Sec. \ref{sec:setup}, we also define $Z^{z}$ in the doubled Hilbert space, 
\begin{align*}
\begin{split}
Z^z =  \text{\fineq{
\draw[ thick] (0,0) -- (1, 0);
\draw[fill=bertinigreen, thick]  (1,0) circle  (0.15);
\node at (1.5,0) {$z$};
}
},
\end{split}
\end{align*}
such that the diagram represents $|0\rangle \,\hat\otimes\, |0\rangle + |1\rangle \,\hat\otimes\, |1\rangle $ if $z =0$ or $|0\rangle \,\hat\otimes\, |0\rangle - |1\rangle \,\hat\otimes\, |1\rangle $ if $z=1$. Furthermore, the following algebraic property of CNOT is essential for obtaining Eq. (\ref{eq:Z_paulistring}),
\begin{align}
 \fineq{ 
 \text{
 \tsfmatV[0][-0.5][r][1][][][blue50]
\draw[fill=bertinigreen, thick]  (-0.5,0) circle  (0.15);
\draw[fill=bertinigreen, thick]  (0.5,0) circle  (0.15);
\node at (-0.5,-0.5 ) {$a$};
\node at (0.5, -0.5 ) {$b$};
} 
}
=
\fineq{
\text{
\draw[ thick] (-0.5,0) -- (-0.5, 1);
\draw[ thick] (0.5, 0 ) -- (0.5,1);
\draw[fill=bertinigreen, thick]  (-0.5,0) circle  (0.15);
\draw[fill=bertinigreen, thick]  (0.5,0) circle  (0.15);
\node at (-0.5,-0.5) {$a \oplus b $};
\node at (0.5,-0.5) {$b$};
}
}
\label{eq:Z_pull_through}
.
\end{align}
It means in particular that if $O$ is a $Z$ string then so is $\tilde O$.
We apply Eq. (\ref{eq:Z_pull_through}) repeatedly to the left hand side of Eq. (\ref{eq:Z_paulistring}), finding

\begin{align}
\begin{split}
\U^{(0)}_{\text{SQC} } \left( \bigotimes_{i = 0}^{L-1} Z^{z_{i}} \right) 
=
\text{
\fineq[-0.8ex][1][1]{
\tsfmatV[-4][3][r][1][][][blue50][topright]
\tsfmatV[-2][1][r][1][][][blue50][topright]
\tsfmatV[-1][0][r][1][][][blue50][topright]
\draw[black,dashed] (-2.75,2.75) -- (-3.25,3.25);
\draw[fill=bertinigreen, thick]  (-0.5,0.5) circle  (0.15);
\draw[fill=bertinigreen, thick]  (-1.5,0.5) circle  (0.15);
\draw[fill=bertinigreen, thick]  (-2.5,1.5) circle  (0.15);
\draw[fill=bertinigreen, thick]  (-4.5,3.5) circle  (0.15);
\node at (-0.5,0) {$z_{L-1}$};
\node at (-1.5,0) {$z_{L-2}$};
\node at (-2.5,1) {$z_{L-3}$};
\node at (-4.5,3) {$z_0$};
}
} 
=
\text{
\fineq[-0.8ex][1][1]{
\draw[black,dashed] (-3,2.5) -- (-4.,2.5);
\draw[ thick] (-0.5,2.5) -- (-0.5,3.25);
\draw[ thick] (-1.5,2.5) -- (-1.5,3.25);
\draw[ thick] (-2.5,2.5) -- (-2.5,3.25);
\draw[ thick] (-4.5,2.5) -- (-4.5,3.25);
\draw[fill=bertinigreen, thick]  (-0.5,2.5) circle  (0.15);
\draw[fill=bertinigreen, thick]  (-1.5,2.5) circle  (0.15);
\draw[fill=bertinigreen, thick]  (-2.5,2.5) circle  (0.15);
\draw[fill=bertinigreen, thick]  (-4.5,2.5) circle  (0.15);
\node at (-0.5,2) {$\tilde z_{L-1}$};
\node at (-1.5,2) {$\tilde z_{L-2}$};
\node at (-2.5,2) {$\tilde z_{L-3}$};
\node at (-4.5,2) {$\tilde z_{0}$};
}
}
\end{split}
\label{eq:UZz}
\end{align}
with $\tilde{z}_i =  \oplus_{j=i}^{L-1}z_j$.  This is Eq.~\eqref{eq:Z_paulistring}, as required.

A similar result holds for the $X$ Pauli-string, because $Z$ and $X$ are related by a Hadamard matrix. Using the above diagrammatic approach, one arrives at the following 
\begin{equation}
      \U^{(0)}_{\text{SQC} } \left( \bigotimes_{i = 0}^{L-1} X^{z_{i}} \right)  = 
      \bigotimes_{i = 0}^{L-1} X^{\oplus_{j = 0 }^{ i }z_{j}}.
\end{equation}
It follows that if $O$ is an $X$ Pauli string then so is $\tilde{O}$, hence $\langle O \rangle_s$ is zero in that case.  Among Pauli strings, only $Z$ strings have finite expectations in Eq.~\eqref{eq:observable_CNOT}.
To calculate the expectations in this case we use Eq.~\eqref{eq:U-tri-sqc} to express $\U^{\triangleleft}$ in terms of the $\U^{(k)}_{\rm SQC}$'s, and use Eq.~\eqref{eq:UZz}, as we now discuss.

\subsection{$1$-point correlation functions}

\begin{figure}[t!]
    \centering
    \includegraphics[width=0.5\columnwidth]{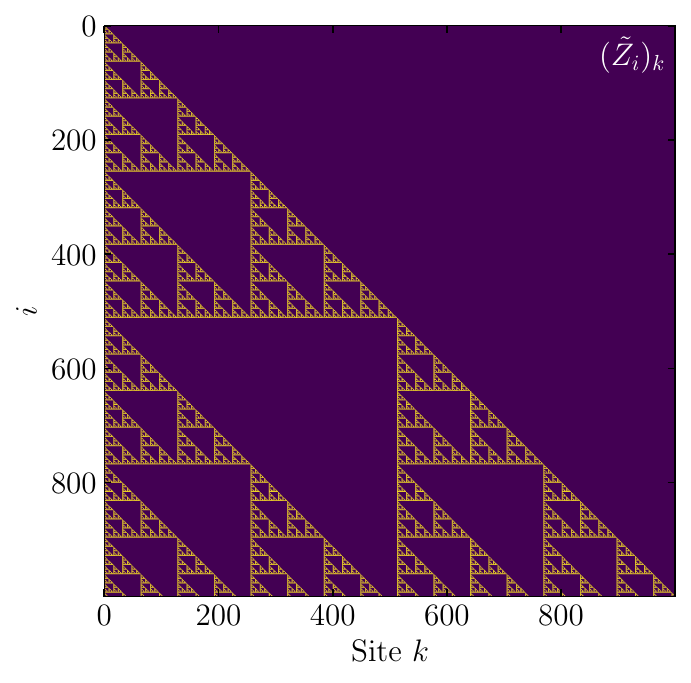}
    \caption{The Pauli-string structure of $\tilde{Z_{i}}$ for $L = 1000$; The rows plot the component $k$ of string $\tilde{Z_{i}}$, where $Z$($\mathbf{1}$) is shown in yellow(purple). The emerged Sierpiński triangle is understood using Eq. (\ref{eq:Z_paulistring_1}). }
    \label{fig:SM1}
\end{figure}

The simplest non-trivial expectation value for a $Z$ Pauli string is $\langle Z_i \rangle_s$, which we identify as a one-point correlation function. We now show that this correlation function 
 has a \textit{Sierpiński triangle} structure (see Fig. \ref{fig:SM1}), in particular
\begin{equation}
    \tilde{Z}_i =  \U^{\triangleleft} ( Z_i) = \prod_{k = 0}^{i } Z_{k}^{ \beta_{i,k} }, 
\label{eq:Z_paulistring_1}
\end{equation}
where  
\beq
\beta_{i,k}=\binom{i}{k} \text{ mod } 2
\eeq 
is the binomial coefficient mod 2.
Note that $\tilde{Z}_i$ is a $Z$ Pauli string where all the operators to the right of site $i$ are identities.  
This means in particular that while the objects in Eq.~\eqref{eq:Z_paulistring_1} are $L$-qubit operators, the equation holds for all $L(> i)$.

A special case of Eq.~\eqref{eq:Z_paulistring} is that  
\beq
{\cal U}_{\text{SQC}}^{(0)}( Z_i ) = \prod_{p=0}^i Z_p
\label{eq:expand_i}
\eeq
From \eqref{eq:def-USQC} the operator ${\cal U}_{\text{SQC}}^{(k)}$ acts as the identity on the first $k$ qubits.
Repeating the argument leading to Eq.~\eqref{eq:Z_paulistring} yields the more general result
\beq
{\cal U}_{\text{SQC}}^{(k)}( Z_i ) = \begin{cases}  
\prod_{p=k}^i Z_p & i\geq k \\ Z_i & i<k
\label{eq:reduce_ik}
\end{cases}
\eeq

Using these results with Eq.~\eqref{eq:U-tri-sqc} and unitarity of ${\cal U}_{\text{SQC}}^{(k)}$ we have
\begin{align} 
\tilde{Z}_i & = {\cal U}_{\text{SQC}}^{(L-2)} \cdots {\cal U}_{\text{SQC}}^{(1)} \, {\cal U}_{\text{SQC}}^{(0)}( Z_i ) 
\nonumber\\
& =  {\cal U}_{\text{SQC}}^{(L-2)} \cdots {\cal U}_{\text{SQC}}^{(1)} \left(  \prod_{p=0}^{i} Z_p \right) 
\nonumber\\
& =\prod_{p=0}^{i} {\cal U}_{\text{SQC}}^{(L-2)} \cdots {\cal U}_{\text{SQC}}^{(1)} \left(  Z_p \right)  
\end{align}
For $i=0$ this gives $\tilde{Z}_0=Z_0$ [by Eq.~\eqref{eq:reduce_ik}].  For $i>0$ we have by Eq.~\eqref{eq:reduce_ik}
\begin{align} 
\tilde{Z}_i  & =  Z_0 \prod_{p=1}^{i} {\cal U}_{\text{SQC}}^{(L-2)} \cdots {\cal U}_{\text{SQC}}^{(1)} \left(  Z_p \right)  
\label{eq:Zi1} 
\end{align}

The key step is to observe [by the properties of ${\cal U}_{\text{SQC}}^{(k)}$ and ${\cal U}^\triangleleft$] that ${\cal U}_{\text{SQC}}^{(L-2)} \cdots {\cal U}_{\text{SQC}}^{(1)} \left(  Z_p \right) = \mathbb{1}\otimes \tilde Z_{p-1}$ (here $\tilde Z_{p-1}$ is interpreted as an operator on $L-1$ qubits, this is consistent because $1\leq p<L$).  Hence
\beq
\tilde{Z}_i  =  Z \otimes \prod_{p=1}^{i} \tilde Z_{p-1} 
\label{eq:Zi2}
\eeq

We now show Eq.~\eqref{eq:Z_paulistring_1} by induction on $i$.  We have shown Eq.~\eqref{eq:Z_paulistring_1} for $i=0$.  Fix some $i$ and suppose that Eq.~\eqref{eq:Z_paulistring_1} holds for all $p< i$.  Then by Eq.~\eqref{eq:Zi2} we have
\begin{align}
\tilde{Z}_i  &=  Z \otimes \prod_{p=1}^{i} \prod_{k = 0}^{p-1} Z_{k}^{ \beta_{p-1,k} }
\nonumber\\
& = Z_0 \prod_{p=1}^{i} \prod_{k = 0}^{p-1} Z_{k+1}^{ \beta_{p-1,k} }
\nonumber\\
& = Z_0 \prod_{k=0}^{i-1} (Z_{k+1})^{\sum_{p=k+1}^{i} \beta_{p-1,k} }
\nonumber\\
& = Z_0 \prod_{k=0}^{i-1} Z_{k+1}^{\beta_{i,k+1} }
\end{align}
where the last line uses Fermat's combinatorial identity $\sum_{a=k}^{c} \binom{a}{k} = \binom{c+1}{k+1}$.
The last result is equivalent to Eq.~\eqref{eq:Z_paulistring_1}, so this concludes the inductive proof.

Using this result with Eqs.~(\ref{eq:observable_CNOT},\ref{eq:Obs_Z_string}), we obtain
\begin{equation}
    \langle Z_{i} \rangle = \left( \frac{ e^{s} - e^{-s} }{ e^{s}  + e^{-s}}\right)^{V_i},
    \label{eq:1pt-s}
\end{equation}
where 
\beq
V_i =  \sum_{k=0}^i \beta_{i,k} \; .
\eeq
is the number of $Z$ operators in the Pauli string $\tilde Z_i$, which is the same as the number of odd numbers in the $i$th row of Sierpiński's triangle.  Hence
Eq.~\eqref{eq:1pt-s} is equivalent to Eq. (\ref{eq:Z}) of the main text.%

\subsection{$2-$point correlation functions}

The two-point correlation function is 
\beq
\langle Z_i Z_j \rangle_s =  \Tr[ {\cal U}^\triangleleft(\tilde{Z}_i) \,{\cal U}^\triangleleft(\tilde{Z}_j) m_{s} ^{\otimes L} ] e^{-L\theta(s)}
\label{eq:ZZ-tr}
\eeq
We consider $i\leq j$ without loss of generality.  Using Eq. (\ref{eq:Z_paulistring_1}) and that $Z^2=\mathbb{1}$ we find
\begin{equation}
      \U^{\triangleleft} ( Z_i ) \,\U^{\triangleleft} (Z_j) = \bigotimes_{k = 0}^{i } Z^{  \beta_{i,k} \oplus  \beta_{j,k} }. 
\label{eq:Z_paulistring_2}
\end{equation}
Using this with Eq.~\eqref{eq:ZZ-tr} yields 
\beq
\langle Z_i Z_j \rangle_s  =  \left( \frac{ e^{s} - e^{-s} }{ e^{s}  + e^{-s}}\right)^{W_{ij}} 
\eeq
where (recall $i\leq j$)
\beq
W_{i,j} = \sum_{k=0}^i (\beta_{i,k} \oplus \beta_{j,k} ) + \!\!\!\sum_{k=i+1}^j \!\!\!\beta_{j,k}
\eeq
is the number of $Z$ operators in the Pauli string ${\cal U}^{\triangleleft}(Z_i Z_j)$.  This is Eq. (\ref{eq:ZZ}) of the main text.

Introducing the connected correlation function, we have
\begin{align}
   C(i,j) &
   \coloneq
    \langle Z_{i}Z_{j} \rangle_s - \langle Z_{i} \rangle_s \langle Z_{j} \rangle_s
   \nonumber\\
    &=
    \left( \frac{ e^{s} - e^{-s} }{ e^{s}  + e^{-s}}\right)^{W_{ij}} - \left( \frac{ e^{s} - e^{-s} }{ e^{s}  + e^{-s}}\right)^{V_i} \left( \frac{ e^{s} - e^{-s} }{ e^{s}  + e^{-s}}\right)^{V_j} \; .
\label{eq:Z_paulistring_c}
\end{align}

We now demonstrate that there are long-range correlations in Eq.~(\ref{eq:Z_paulistring_c}) by exploring features of the Sierpiński triangle. We make use of Lucas' theorem, which states that $\binom{i}{k} \mod 2$ is $0$ if and only if at least one digit of the binary representation of $k$ is greater than the corresponding digit for $i$. Hence for integer $m$, $\binom{2^{m}-1}{k} \text{ mod } 2$ is $1$ for all $k$; also $\binom{2^{m}}{k} \text{ mod } 2$ is $1$ if and only if  $k=0,2^m$.
It follows that
\beq
    V_{2^m-1} = 2^m , \qquad 
    V_{2^m} = 2
\eeq
Similarly, for two-point correlations:
\begin{align*}
    W_{2^m, 2^{m'}} &= 2 \\ 
    W_{2^m - 1, 2^{m'} - 1} &= 2^{m'} - 2^m \\
    W_{2^m , 2^{m} - 1} &= 2^{m} .
\end{align*}
Combining these results yields Eq. (24) of the main text.

A particularly interesting case is $i = 2^m, j = 2^{m'}$, where $W_{i,j}$, $V_i$ and $V_j$ are all equal to $2$, independent of $m$ and $m'$. This means the connected $2$-point function Eq. (\ref{eq:Z_paulistring_c}) is the same and non-zero for any $m$ and $m'$. Since $|2^m - 2^{m'}|$ can be arbitrarily large, these are infinite-ranged correlations.

\subsection{Averaged correlations}

The correlation function $C(i,j)$ has a complicated dependence on $i,j$; the general trend is that larger distances $|i-j|$ correspond to weaker correlations, but there are some special points where the correlation is finite even for very large distances.

To reveal the general trend we consider averages over spatial windows: for any vector $f=(f_0,f_1,\dots)$ we write 
\beq
[f]_a^b = \frac{1}{b-a} \sum_{j=a}^{b-1} f_j
\eeq
where the window is $a\leq j<b$.
Also note from the recursive structure of the Sierpiński's triangle that for integers $j,m$ with $j<2^m$:
\be
\label{eq:basicrelation}
V_{j+2^{m}} = 2 V_j.
\ee
 Considering the vector $(V_0^k,V_1^k,\dots)$ where we take the $k$th power of each $V_i$, this implies
\be
\begin{aligned}
[V^k]_0^{2^{m{+1}}} &= \frac{1}{2^{m+1}} \sum_{j=0}^{2^{m+1}-1} V_j^k =  \frac{1}{2^{m+1}} \left(\sum_{j=0}^{2^{m}-1} V_j^k + \sum_{j=0}^{2^{m}-1} V_{j+2^m}^k \right)\\
& =\left(\frac{1+2^k}{2}\right) [V^k]_0^{2^{m}}
\end{aligned}
\ee
combining this relation with $[V^k]_0^{1}=1$ we find 
\be
[{V^k}]_0^{2^{m}} = \left(\frac{1+2^k}{2}\right)^m = 2^{m(d_{\frak f}(k)-1)}\,. 
\label{eq:ind-Vk}
\ee
with 
\be
d_{\frak f}(k)= \frac{\log(1+2^k)}{\log 2}\,. 
\ee

\subsubsection{One-point function}

Define $q(s)=\ln \tanh(s)$ as in the main text, we consider $s>0$ so $q<0$.   We have 
\beq
\langle Z_i\rangle_s = \exp[ q(s) V_i ]
\eeq
(For $s<0$ then $e^q$ is real and negative, and $q$ is complex.)
Now we compute the one-point correlation function averaged over the window from $0$ to $2^{m}$ as 
\be
\label{eq:expZ}
[\expval{Z}_s]_0^{2^{m}} = \sum_{k=0}^\infty \frac{q^k}{k!} [V^k]_0^{2^{m}} = \sum_{k=0}^\infty \frac{q^k}{k!} \left(\frac{1+2^k}{2}\right)^{m}= \frac{1}{2^m}\sum_{j=0}^m \binom{m}{j} e^{2^j q},
\ee
where the last form follows from expanding the binomial and resumming the exponential. This expression can be bounded as follows
\be
\frac{1}{2^m} e^{2q}\leq [\expval{Z}_s]_0^{2^{m-1}} \leq   \frac{1}{2^m} \binom{m}{m/2} \frac{e^{q}}{1-e^q}, 
\label{eq:bounds}
\ee
where we took for simplicity $m$ to be even.  The lower bound is obtained by only taking the first term in the sum (they are all positive) and the upper bound by setting all binomial coefficients equal to the largest one and using 
\be
\sum_{j=0}^m  e^{2^j q} \leq \sum_{k=1}^{2^m} e^{k q} \leq \frac{e^{q}}{1-e^q}\,. 
\ee
Note that the lower bound in Eq.~\eqref{eq:bounds} scales as $2^{-m}$ and since 
\be
 \frac{1}{2^m} \binom{m}{m/2} = \sqrt\frac{2}{m \pi} + O\left (\frac{1}{m^{3/2}}\right), 
\ee
the upper bound scales like $m^{-1/2}$. 

Finally we consider a sequence of system sizes that are powers of two ($L=2^p$) and average the one-point function over the whole system to bound the averaged correlation
\beq
C_1 < [\langle Z\rangle_s]_0^{L} < C_2, \qquad C_1 \sim L^{-1}, \qquad C_2 \sim (\log L)^{-1/2}
\eeq
This upper bound shows that $ [\langle Z\rangle_s]_0^{L}$ decays with $L$; the lower bound shows that this decay is no faster than a power law.

\subsubsection{Two-point function}

The two-point function can be treated in a similar way.  We consider the illustrative case $i=2^m \leq j$ for which
\be
\expval{Z_i Z_j}_s = \expval{Z_j}_s \exp[- 2 q \beta_{j,i}]
\ee
where we used that $\beta_{j,0}=1$. It is also useful to note that Lucas's theorem implies 
\be
\beta_{j,2^m} =j_m, 
\ee
where $j_m$ is the $m$-th coefficient of the expansion of $j$ in base 2. This means that we can write the following window-averaged correlation:
\be
[\expval{Z_{2^m} Z_j}_s]_{2^n}^{2^{n+1}} = \frac{1}{2^n} \left(\sum_{k=0}^{2^{n-m-1}-1} \sum_{a=0}^{2^{m}-1} \expval{Z_{2^n+a+2^{m}(2k+1)}}_s +  e^{- 2 q} \sum_{k=0}^{2^{n-m-1}-1} \sum_{a=0}^{2^{m}-1} \expval{Z_{2^n+a+2^{m+1}k}}_s \right). 
\eeq

To simplify this expression, we  introduce the {($m$-dependent)} parity-split window averages of $f$ 
\be
[f']_{2^n}^{2^{n+1}} : = \frac{1}{2^n} \sum_{k=0}^{2^{n-m-1}-1} \sum_{a=0}^{2^{m}-1} f_{a+(2k+1)2^{m}+2^n},\qquad \qquad \mathbb [f'']_{2^n}^{2^{n+1}} : = \frac{1}{2^n} \sum_{k=0}^{2^{n-m-1}-1} \sum_{a=0}^{2^{m}-1} f_{a+(2k)2^{m}+2^n},
\ee
such that
\be
[f]_{2^n}^{2^{n+1}} = [f']_{2^n}^{2^{n+1}}+[f'']_{2^n}^{2^{n+1}}\,.
\ee

Proceeding inductively by analogy with the derivation of Eq.~\eqref{eq:ind-Vk} we find
\be
[{V^k}']_{2^n}^{2^{n+1}}  =\left(\frac{2^{2k}}{1+2^k}\right) \left(\frac{1+2^k}{2}\right)^n,\quad [{V^k}'']_{2^n}^{2^{n+1}} =\left(\frac{2^{k}}{1+2^k}\right) \left(\frac{1+2^k}{2}\right)^n,\quad [{V^k}]_{2^n}^{2^{n+1}} = 2^k \left(\frac{1+2^k}{2}\right)^n.
\ee
Despite the dependence of the window-split average on $m$, 
it is important here that all these objects are independent of this parameter, which is due to Eq.~\eqref{eq:basicrelation}.
Using the results above we then have 
\be
\begin{aligned}
[\expval{Z_{2^m} Z}_s]_{2^n}^{2^{n+1}} & = \sum_{p=0}^\infty \frac{q^p}{p!} \left(\frac{2^{2p}}{1+2^p}\right) \left(\frac{1+2^p}{2}\right)^n+  e^{- 2 q} \sum_{p=0}^\infty \frac{q^p}{p!} \left(\frac{2^{p}}{1+2^p}\right) \left(\frac{1+2^p}{2}\right)^n\\
&=\frac{1}{2^n}\sum_{j=0}^{n-1} \binom{n-1}{j} e^{2^{j+2} q}+ \frac{e^{-2q}}{2^n}\sum_{j=0}^{n-1} \binom{n-1}{j} e^{2^{j+1} q}\,,
\end{aligned}
\ee
and analogously [cf. Eq.~\ref{eq:expZ}]
\be
[\expval{Z}_s]_{2^n}^{2^{n+1}} = \sum_{p=0}^\infty \frac{(2q)^p}{p!}\left(\frac{1+2^p}{2}\right)^n = \frac{1}{2^n}\sum_{j=0}^n \binom{n}{j} e^{2^{j+1} q}. 
\ee

Finally putting all together, we obtain an expression for the window-averaged two-point function and $i=2^m$
\be
\begin{aligned}
[\expval{Z_{i} Z}_s-\expval{Z_{i}}_s\expval{Z}_s]_{2^n}^{2^{n+1}} &=\frac{1}{2^n}\sum_{j=0}^{n-1} \binom{n-1}{j} 
\left[ e^{2^{j+2} q} - 2 \sinh(2q)e^{2^{j+1} q}\right]+e^{2 q} \frac{e^{2^{n+1} q}}{2^n}\,. 
\end{aligned}
\ee
Similar to Eq.~\eqref{eq:expZ}, this expression can be bounded to show that the correlations decay with $n$ (no faster than algebraically, note $\sinh(2q)<0$), see the main text for a discussion.  The fact that the RHS is independent of $m$ shows that the decay of these correlations is controlled by the distance of the rightmost site from the origin, instead of by the distance between sites $i,j$.

\end{document}